\documentclass [a4paper]{article}

\usepackage{jcappub}
\usepackage{graphicx}
\usepackage[sort&compress]{natbib}

\newcommand{\goo}{{\rm goo}}

\begin{document}

\subheader{CERN-PH-TH/2011-156, LAPTH-023/11}

\title{Dark goo: Bulk viscosity as an alternative to dark energy}
\author[a]{Jean-Sebastien Gagnon,}
\author[b]{Julien Lesgourgues}

\affiliation[a]{Technische Universit\"{a}t Darmstadt,\\ Schlossgartenstrasse 2, 64289, Darmstadt, Germany}
\affiliation[b]{\'{E}cole Polytechnique F\'{e}d\'{e}rale de Lausanne, CH-1015, Lausanne, Switzerland;\\
CERN, Theory Division, CH-1211 Geneva 23, Switzerland;\\
LAPTh (CNRS - Universit\'e de Savoie), BP 110, F-74941 Annecy-le-Vieux Cedex,
France.
}

\emailAdd{jean-sebastien.gagnon@physik.tu-darmstadt.de}
\emailAdd{julien.lesgourgues@cern.ch}

\abstract{We present a simple (microscopic) model in which bulk
  viscosity plays a role in explaining the present acceleration of the
  universe.  The effect of bulk viscosity on the Friedmann equations
  is to turn the pressure into an ``effective'' pressure containing
  the bulk viscosity.  For a sufficiently large bulk viscosity, the
  effective pressure becomes negative and could mimic a dark energy
  equation of state. Our microscopic model includes self-interacting
  spin-zero particles (for which the bulk viscosity is known) that are
  added to the usual energy content of the universe. We study both
  background equations and linear perturbations in this model.  We
  show that a dark energy behavior is obtained for reasonable values
  of the two parameters of the model (i.e. the mass and coupling of
  the spin-zero particles) and that linear perturbations are
  well-behaved.  There is no apparent fine tuning involved.  We also
  discuss the conditions under which hydrodynamics holds, in
  particular that the spin-zero particles must be in local equilibrium
  today for viscous effects to be important.}


\maketitle
\flushbottom

\section{Introduction}

Experimental evidences --- such as type IA supernovae luminosity~\cite{Riess_etal_1998,Perlmutter_etal_1998}, Cosmic Microwave Background 
(CMB) peaks~\cite{WMAP7_2010}, Baryon Acoustic Oscillations (BAO)~\cite{Eisenstein_etal_2005} and Large Scale Structures (LSS)~\cite{Tegmark_etal_2006} --- point toward a universe made of 70\% ``dark energy'' (for reviews see \cite{Amendola_Tsujikawa_2010,Frieman_etal_2008}).  This dark energy is responsible for the observed acceleration of type IA supernovae and is thus repulsive (i.e. it has negative pressure).  Explaining this acceleration has become a important challenge for theoretical physics.

There exist many candidate mechanisms/models to explain this acceleration (for reviews see \cite{Li_Wang_2011,Amendola_Tsujikawa_2010,Caldwell_Kamionkowski_2009,Brax_2009}).  The most famous one is to add a cosmological constant to Einstein's equations.  Although being extremely economical, this solution suffers from severe fine tuning problems. This lead theorists to investigate alternative scenarios, such as modified matter content models (e.g. quintessence~\cite{Caldwell_etal_1998}), modified gravity models (e.g. scalar-tensor theories~\cite{Uzan_1999,Amendola_1999,Chiba_1999}) and models relying on the statistical distribution of matter in the universe (e.g. backreaction~\cite{Rasanen_2003,Kolb_etal_2005,Buchert_2007} or voids~\cite{Tomita_1999,Celerier_1999}).  The present status is that none of these models offer a satisfactory solution to the dark energy problem because they either suffer from severe fine tuning problems, lead to instabilities, are ruled out or are awaiting further analysis.

In this paper we present an alternative (microscopic) model of dark energy that includes the effects of 
bulk viscosity.  In an expanding system, relaxation processes associated with bulk viscosity effectively 
reduce the pressure as compared to the value prescribed by the equation of state.  For a sufficiently 
large bulk viscosity, the effective pressure becomes negative and could mimic a dark energy behavior.  

The idea of having the bulk viscosity drive the acceleration of the universe is mentioned in ref.~\cite{Padmanabhan_Chitre_1987}.  Unified Dark Matter models take this idea one step further and use a single (bulk) viscous fluid to explain both dark matter and dark energy.  These studies postulate an 
exotic equation of state for viscous matter of the form $\zeta = \zeta_{0}\rho^{m}$ (where $\zeta_{0},m$ are parameters)~\cite{Murphy_1973,Belinskii_Khalatnikov_1975} and investigate its effect on the evolution of the universe~\cite{Fabris_etal_2005,Colistete_etal_2007,Li_Barrow_2009,Velten_Schwarz_2011}.  It is interesting to note that the Chaplygin gas model~\cite{Kamenshchik_etal_2001} is a special case of these bulk viscous models.  The conclusion is that it is difficult to explain both dark matter and dark energy using a single viscous fluid~\cite{Li_Barrow_2009,Amendola_Tsujikawa_2010,Velten_Schwarz_2011}.  The reason for this is that bulk viscosity becomes important at low redshifts (in order to have a negative pressure) and density fluctuations in the viscous fluid are quickly damped.  The resulting decay of the gravitational potential (obeying the Poisson equation) has a dramatic effect on structure formation and leads to an integrated Sachs-Wolfe effect on CMB anisotropies that exceeds observational bounds~\cite{Li_Barrow_2009,Velten_Schwarz_2011}.

We take a different, more microscopic approach here (for a different microscopic model with bulk viscosity induced by dark matter annihilation, see references~\cite{Wilson_etal_2006,Mathews_etal_2008}).  Our model consists in adding a new component to the energy content of the universe coming from a self-interacting scalar field.  Thanks to recent progress in thermal field theory, the bulk viscosity of scalar theories has been computed from first principles~\cite{Jeon_1995}.  The resulting functional form depends on physical properties of the scalar particle (mass and self-coupling) and is very different from previous viscous models.  We study the consequences of this new scalar component (including bulk viscosity) on the evolution of the universe.  We emphasize that our model aims at explaining dark energy only	and the arguments against Unified Dark Matter models with bulk viscosity do not apply here.  Indeed, the scalar component of our model is added on top of the usual Cold Dark Matter (CDM) component and thus the late damping of fluctuations in the viscous fluid just results in dark energy being very homogeneous today without affecting structure formation or the CMB.

The rest of the paper is organized as follows.  In section~\ref{sec:Theoretical_concepts} we present 
some theoretical concepts related to hydrodynamics and bulk viscosity.  We also present our model and 
the conditions under which it is valid.  Sections~\ref{sec:Background_evolution} and~\ref{sec:Linear_perturbations} are dedicated to the study of background equations and linear perturbations respectively.  In each case we derive the modifications to the equations due to bulk 
viscosity and show the results of simulations done with the Cosmic Linear Anisotropy Solving System\footnote{\tt http://class-code.net} 
({\tt CLASS}) code~\cite{Lesgourgues_2011,Blas_etal_2011}.  The compatibility with observations of our model is presented in section~\ref{sec:Compatibility_observations}.  We then conclude in Section~\ref{sec:Conclusion}.

\section{Theoretical concepts and presentation of the model}
\label{sec:Theoretical_concepts}

\label{sec:Bulk_viscosity}

Hydrodynamics is an effective theory that describes the evolution of ``fluid cells'', where the equations of 
motion are given by exact local conservation of energy-momentum (we implicitly assume that there are no 
other conserved charges in the following).  A fluid cell is a macroscopic average (of size $l$) over many 
fundamental fluid constituents.  For a perfect fluid, macroscopic quantities such as pressure $p$, energy 
density $\rho$ and velocity $v$ are smooth inside a fluid cell and local thermal equilibrium is 
maintained.  For a viscous fluid, microscopic interactions make $p$, $\rho$ and $v$ vary appreciably 
over a mean free path and local thermal equilibrium is not maintained inside a fluid cell.  The effect of 
dissipation due to microscopic interactions is characterized by transport coefficients (shear and bulk 
viscosities).  In the following we focus on bulk viscosity since we are interested in the universe and there 
is no shear in an expanding system.

Viscous corrections to the homogeneous evolution of the universe are discussed in great details in ref.~\cite{Weinberg_1972}.  Let us repeat the salient points.  The energy-momentum tensor for a perfect fluid 
is given by:
\begin{eqnarray}
\label{eq:Tmunu_perfect}
T_{\rm perfect}^{\mu\nu} & = & pg^{\mu\nu} + (p + \rho)U^{\mu}U^{\nu},
\end{eqnarray}
where $g^{\mu\nu}$ is the metric tensor and $U^{\mu}$ is the velocity four-vector (normalized such that $U_{\mu}U^{\mu} = -1$).  In a Friedmann-Robertson-Walker universe with metric $g^{\mu\nu} = \mbox{diag}(-1,a,a,a)$, the usual procedure leads to the Friedmann equation:
\begin{eqnarray}
\label{eq:Friedmann_perfect}
\frac{\dot{a}^{2}}{a^{2}} + \frac{k}{a^{2}} & = & \frac{8\pi G}{3}\rho, 
\end{eqnarray}
and the energy conservation equation:
\begin{eqnarray}
\label{eq:Energy_conservation_perfect}
\dot{\rho} & = & -\frac{3\dot{a}}{a}\left(\rho + p \right),
\end{eqnarray}
where $a$ is the scale factor, $k$ is the spatial curvature and dots represent derivatives with respect to 
time.  To include viscous corrections, we need the expression for the viscous energy-momentum tensor.  
Transport coefficients characterize linear deviations away from equilibrium due to thermodynamic forces 
(gradients).  In a viscous fluid, momentum flows between neighboring fluid cells due to microscopic 
interactions and is thus driven by velocity gradients.  The most general energy-momentum tensor 
linear in velocity gradients that satisfies the second law of thermodynamics for all fluid configurations is~\cite{Weinberg_1972}:
\begin{eqnarray}
\label{eq:Tmunu_viscous_1}
T_{\rm viscous}^{\mu\nu} & = & T_{\rm perfect}^{\mu\nu} -\zeta \left(g^{\mu\nu} + U^{\mu}U^{\nu}
\right)D_{\gamma}U^{\gamma},
\end{eqnarray}
where $\zeta$ is the bulk viscosity and $D_{\gamma}$ is the covariant derivative.  In a Friedmann-Robertson-Walker universe, we have $D_{\gamma}U^{\gamma} = 3\dot{a}/a$ and we re-write the viscous energy-momentum tensor as:
\begin{eqnarray}
\label{eq:Tmunu_viscous_2}
T_{\rm viscous}^{\mu\nu} & = & \rho U^{\mu}U^{\nu} + \left(p -3\zeta \frac{\dot{a}}{a}\right)\left(g^{\mu\nu} 
+ U^{\mu}U^{\nu}\right) \nonumber \\
												 & \equiv & p_{\rm eff}g^{\mu\nu} + (p_{\rm 
eff} + \rho)U^{\mu}U^{\nu}.
\end{eqnarray}
From eqs~\ref{eq:Tmunu_perfect} and \ref{eq:Tmunu_viscous_2}, we see that the effect of bulk viscosity 
is to change the pressure $p$ to an effective pressure $p_{\rm eff} = p -3\zeta \dot{a}/a$.  The physical 
interpretation is clear.  An expanding fluid leaves its equilibrium state; the energy density decreases and 
the pressure also decreases.  In the absence of bulk viscosity, the fluid relaxes instantaneously and 
pressure and density are related by the equation of state.  Bulk viscosity dampens this behavior by 
introducing a finite relaxation timescale, hence producing a shift between the equation of state pressure 
and the true pressure.

The background evolution equations are obtained by replacing $p$ by $p_{\rm eff}$ in the energy 
conservation equation~\ref{eq:Energy_conservation_perfect} (the Friedmann equation~
\ref{eq:Friedmann_perfect} remains unchanged).  We note that for a large enough $\zeta$, the effective 
pressure becomes negative and could mimic a dark energy behavior.  The bulk viscosity coefficient $\zeta$ depends on microscopic interactions and must be computed from a more fundamental theory.

\subsection{Presentation of the model}
\label{sec:Presentation_model}

The background equations~\ref{eq:Friedmann_perfect} and~\ref{eq:Energy_conservation_perfect} are 
effective equations that do not rely on microscopic physics, except through the bulk viscosity parameter.  
We thus need some microscopic physics to ``personify'' the bulk viscosity.  For reasons we discuss in 
section~\ref{sec:Validity_hydrodynamics}, no Standard Model particles can produce a bulk viscosity in 
the late universe.  We thus need to add a new component to the energy content of the universe in order 
to induce some bulk viscosity in the present universe.

For simplicity, we add a massive scalar field with quartic interactions to the Lagrangian of the Standard 
Model (cubic interactions would change our results in a trivial way and are left out in the following):
\begin{eqnarray}
\label{eq:Lagrangian}
{\cal L} & = & \frac{1}{2}\partial_{\mu}\phi\partial^{\mu}\phi - \frac{1}{2}m_{0}^{2}\phi^{2} - \frac{\lambda}{4!}\phi^{4}.
\end{eqnarray}
We also make several assumptions.  First we require that the assumptions of hydrodynamics for this 
scalar fluid are satisfied.  The validity of hydrodynamics is crucial for the concept of bulk viscosity to 
make sense.  These assumptions include local thermal equilibrium, small inhomogeneities (gradients) 
and absence of instabilities.  Each of these points are discussed in section~\ref{sec:Validity_hydrodynamics}.  We also assume that the scalar and Standard Model particles are in thermal equilibrium in the early universe and decouple from each other afterward.  After this decoupling time, the temperature of the
self-interacting scalars has a distinct evolution from that of photons and other Standard Model particles in equilibrium with the photons.  Between electron-positron annihilation and today, the photon temperature scales as $T_\gamma \propto a^{-1}$.  We shall see later that the scalar temperature $T_{s}$ scales in the same way at least between the time of scalar-photon decoupling and the non-relativistic transition ($T_s \sim m_0$).  Hence, the ratio $T_s/T_\gamma$ is expected to remain constant over a long time interval, typically from Big Bang Nucleosynthesis until the end of radiation domination.

The model has three free parameters: the mass $m_{0}$, the self-coupling $\lambda$ of the scalar 
particles, and the ratio of scalar to photon temperatures $\epsilon = T_{s}/T_{\gamma}$ after electron-positron annihilation.  Note that $\epsilon \leq 1$ because $(a T_\gamma)$ can increase due entropy releases until electron-positron annihilation, while $(a T_s)$ stays constant (no interaction with the Standard Model particles) after scalar-photon decoupling.  The limit $\epsilon = 1$ correspond to late decoupling not followed by any entropy release and the limit $\epsilon \ll 1$ corresponds to early decoupling with significant entropy releases.

We stress that our model is very different from quintessence models.  Both include a scalar field with a potential, but the range of parameters is different.  In quintessence models, the scalar field is seen as classical and coherent over large scales, and the potential is adjusted in such a way as to obtain slow-roll conditions in the late universe.  In contrast, we are dealing here with a fluid of scalar particles, and we adjust the parameters of our model in order to allow collective (bulk viscous) effects to play a significant role in this imperfect fluid in the present universe.  

We also mention that our model is not a generalization of previous bulk viscous or Chaplygin gas 
models.  The functional form we use for the bulk viscosity (see section~\ref{sec:Pressure_density_viscosity}) is very different from the (generalized) Chaplygin form.  The bulk viscosity for a scalar theory has been computed from first principles and depends on the physical properties (mass and self-coupling) of the scalar particles; no such interpretation is available for the (generalized) Chaplygin gas form.

\subsection{Pressure, density and bulk viscosity of a scalar field}
\label{sec:Pressure_density_viscosity}

We need the pressure, energy density and bulk viscosity of a scalar field to integrate eqs.~\ref{eq:Friedmann_perfect} and \ref{eq:Energy_conservation_perfect}.  The pressure of a gas of massive scalar particles (with vanishing chemical potential) is given by (e.g. \cite{Gale_Kapusta_2006}):
\begin{eqnarray}
\label{eq:Pressure_scalar}
p & = & \frac{m_{0}^{2}T_{s}^{2}}{2\pi^{2}}K_{2}\left(\frac{m_{0}}{T_{s}}\right) + O(\lambda T_{s}^{4}),
\end{eqnarray}
where $K_{2}$ is a modified bessel function of the second kind.  In the following we neglect $O(\lambda T_{s}^{4})$ since we expect bulk viscous effect to dominate at low temperature.  Using the thermodynamic relation $\rho = T^{2} \partial \left(p/T\right)/\partial T$, we obtain the corresponding energy density:
\begin{eqnarray}
\label{eq:Energy_density_scalar}
\rho & = & \frac{m_{0}^{2}T_{s}}{2\pi^{2}}\left[T_{s} K_{2}\left(\frac{m_{0}}{T_{s}}\right)-\frac{m_{0}}
{2}\left(K_{1}\left(\frac{m_{0}}{T_{s}}\right)+K_{3}\left(\frac{m_{0}}{T_{s}}\right) \right)\right].
\end{eqnarray}

In a remarkable {\em tour de force}, Jeon has computed the leading order bulk viscosity in a scalar 
theory from diagrammatic methods \cite{Jeon_1995}.  The computation involves the resummation of an 
infinite number of diagrams and can only be obtained numerically.  A best fit of the numerical result is:
\begin{eqnarray}
\label{eq:Bulk_viscosity_scalar}
\zeta & = & \left(\frac{\tilde{m}^{4}}{\lambda^{4}m_{\rm th}}\right) \left(\frac{m_{\rm th}}{T_{s}}
\right)^{\kappa_{3}} e^{\kappa_{1}} e^{\kappa_{2} m_{\rm th}/T_{s}},
\end{eqnarray}
where the constants $\kappa_{1}$, $\kappa_{2}$ and $\kappa_{3}$ are 12.13, 1.54 and 1.13, respectively.  The thermal mass (i.e. the sum of the zero-temperature mass plus thermal modifications) is given by $m_{\rm th}^{2} = m_{0}^{2} + \delta m_{\rm th}^{2}$, where the thermal modification is $\delta m_{\rm th}^{2} = \lambda T_{s}^{2}/24$.  The quantity $\tilde{m}^{2} \equiv m_{\rm th}^{2} - (T_{s}/2)(\partial m_{\rm th}^{2}/\partial T_{s})$ gives the breaking of scale invariance in the theory (at lowest order).  More precisely, it is given by $\tilde{m}^{2}  = m_{0}^{2} - \beta(\lambda)T_{s}^{2}/48$ where the beta function is $\beta(\lambda) = 3\lambda^{2}/(16\pi^{2})$.  Note that eq.~\ref{eq:Bulk_viscosity_scalar} is only valid for weak coupling $\lambda < 1$.

We can understand eq.~\ref{eq:Bulk_viscosity_scalar} in the following way.  Bulk viscosity characterizes 
the relaxation of an expanding fluid.  As a result, it is proportional to the mean free time between 
collisions of the fluid constituents.  The mean free time is itself inversely proportional to the cross section 
of the processes responsible for re-establishing equilibrium.  In an expanding system, equilibrium is 
achieved by number changing processes ($2\rightarrow 4$ in $\lambda \phi^{4}$).  This explains the $\lambda^{-4}$ dependence of $\zeta$.  The exponential is due to Boltzmann suppression of final states for these number changing processes.

The bulk viscosity also vanishes in a conformal theory, since in that case dilatation is a symmetry and 
the fluid cannot leave equilibrium.  It is thus proportional to the (square) of the breaking of scale 
invariance (i.e. $\tilde{m}^{4}$).  Consequently, $\zeta$ has a very different coupling constant behavior 
at different temperatures.  At low temperature $\tilde{m}^{4} \sim m_{0}^{4}$ and $\zeta \sim \lambda^{-4}$; at high temperature $\tilde{m}^{4} \sim \beta^{2}$ and $\zeta \sim \lambda^{(\kappa_{3}-1)/2}$, implying that the bulk viscosity is highly suppressed in this regime.

%
%
%
%

\subsection{Validity of the hydrodynamic approximation}
\label{sec:Validity_hydrodynamics}

The crucial point in this model is that the new scalar degree of freedom can be considered as a fluid in 
the dark energy domination epoch.  Otherwise, the concept of bulk viscosity does not make sense.  Thus 
the model must satisfy the assumptions of hydrodynamics.  Let's study each assumption separately.

\begin{description}
	\item[Local equilibrium.] For a hydrodynamic description to be valid, the fluid constituents must 
interact sufficiently often so as to maintain (approximate) local equilibrium (the constituents must not be 
``frozen out'').  More technically, the need for local equilibrium is reflected in the fact that transport 
coefficients are linear deviations away from equilibrium and are computed from equilibrium correlators~\cite{Jeon_1995}.

The criterion to be in equilibrium in an expanding universe is $\Gamma_{\rm scalar}/H \gg 1$, where $\Gamma_{\rm scalar}$ is the rate of processes that are ultimately responsible for the bulk viscosity and 
$H$ is the expansion rate of the universe.  The rate $\Gamma_{\rm scalar}$ is estimated as follows~\cite{Jeon_1995}.  First we have that $\Gamma_{\rm scalar} \sim 1/\tau_{\rm free}$ where $\tau_{\rm free} = n/(dW/dVdt)$ is the mean free time between collisions, $n$ is the density of particles and $dW/dVdt$ is the 
transition rate per unit volume.  The processes that are responsible for bulk viscosity in $\lambda\phi^{4}$ theory are $2\rightarrow 4$ scatterings.  We expect the bulk viscosity to dominate at low 
temperature due to the exponential factor in eq.~\ref{eq:Bulk_viscosity_scalar}.  For low temperatures $T \ll m_{0}$ we have $n = dN/dV \sim O(e^{-m_{0}/T_{s}})$ (Boltzmann suppression) and $dW/dVdt \sim O(\lambda^{4}e^{-4m_{0}/T_{s}})$ (there should be a $m_{0}$ to get the 
units right).  Thus the rate of bulk viscosity processes are:
\begin{eqnarray}
\label{eq:Number_changing_processe_rates}
\Gamma_{\rm scalar} & \sim & \frac{1}{\tau} \;\;\sim\;\; \frac{m_{0}\lambda^{4}e^{-4m_{0}/
T_{s}}}{e^{-m_{0}/T_{s}}} \;\;\sim\;\; m_{0}\lambda^{4}e^{-3m_{0}/T_{s}}.
\end{eqnarray}
The equilibrium criterion becomes:
\begin{eqnarray}
\frac{m_{0}\lambda^{4}e^{-3m_{0}/T_{s}}}{H}\;\; > \;\; 1.
\end{eqnarray}
This criterion must be fulfilled until the present epoch for the model to make sense; this is thus a 
constraint on the parameters of the model.  This also explains why no Standard Model particles produce 
a large bulk viscosity in the late universe, since they all decouple before dark energy domination.

	\item[Small inhomogeneities and gravitational instabilities.] Hydrodynamics is an expansion in 
gradients (see for example eq.~\ref{eq:Tmunu_viscous_1}).  For this expansion to be well defined, 
velocity gradients must be small (small inhomogeneities).  Similarly, clumps of scalar matter may form 
due to gravity.  If these clumps become too large (exponential instabilities), then the fluid breaks apart 
and a hydrodynamic description is not appropriate anymore.

It is difficult to estimate the size of gradients a priori.  For that, a full numerical study of linear 
perturbations is necessary (a dynamical system analysis might also shed some light on these issues, e.g.~\cite{Szydlowski_Hrycyna_2006}).  We perform such a numerical analysis in section~\ref{sec:Linear_perturbations}.  Our results show that linear perturbations are well behaved and no instability develops (see section~\ref{sec:Results_perturbations}).  
 
It is important to emphasize that we are using first order (in gradients) relativistic hydrodynamics.  It has been pointed out in Ref.~\cite{Israel_Stewart_1979} that first order hydrodynamics is acausal and unstable (for certain modes);  to cure these problems, second order hydrodynamics should be used instead (see \cite{Piattella_etal_2011} for a cosmological case study).  A more recent analysis shows that acausality implies unstable behavior~\cite{Denicol_etal_2008,Pu_etal_2009}.  The physical explanation is that an acausal disturbance propagating in a light cone would create a singular behavior at the edge of the light cone, since a light cone cannot be crossed in a covariant theory.  In our case, growing modes are damped out by bulk viscosity and no instability develops.  We conclude that our hydrodynamic equations are well defined and do not produce acausal signal propagation.  It thus seems that the use of linear relativistic hydrodynamics is justified for the modes considered here.

	\item[Thermodynamical instabilities.] This instability is called ``cavitation'' (for a discussion, see e.g. \cite{Brennen_1995}).  In everyday engineering systems, cavitation manifests itself when the pressure of a fluid goes below its saturated vapor pressure.  In such a case, the fluid becomes unstable 
to the formation of bubbles.  The formation of bubbles is akin to the formation of clumps; if there are too 
many bubbles in the fluid, then a hydrodynamic description is not suitable anymore.
	
The process of bubble formation is called nucleation.  There are two types of nucleation: homogeneous 
and heterogeneous.  Homogeneous nucleation happens when a fluid is pure and smooth.  
Heterogeneous nucleation happens when a fluid has impurities and is in contact with boundaries.

In our model, the saturated vapor pressure is the vacuum pressure (i.e. zero).  Thus when the pressure 
becomes negative (required for dark energy domination), the fluid becomes in principle unstable to the 
formation of vacuum bubbles.  In the following we use the language of homogeneous nucleation since 
the fluid is very pure (except maybe for ``gravitational'' impurities) and smooth (no boundaries).

In everyday engineering systems, it is possible to reduce the pressure of a fluid under its saturated vapor 
value without forming bubbles (a bit like a supercooled fluid).  In other words, a normal fluid can 
withstand ``some'' under saturated vapor pressure, it is not automatic that bubbles form.  The physics 
behind that behavior is the following.  Under normal conditions, there is formation of voids between the 
fluid constituents due to thermal fluctuations.  When a fluid is subjected to a pressure that is under the 
saturated vapor pressure, these voids grow and form bubbles.  But there is a counter force to this growth 
due to the potential between fluid constituents.  When the outward pressure is greater than the force 
between constituents, bubbles grow; if not, then the fluid stays as it is.  This property of a fluid is called 
``tensile strength''.  For homogeneous nucleation, tensile strength depends only on the properties of the 
fluid; in heterogeneous nucleation, this tensile strength is strongly dependent on impurities and 
boundary conditions.

Let's denote $p_{b\;\rm max}$ the maximum pressure our scalar fluid can sustain without breaking apart 
(i.e. its tensile strength) and $p_{\rm DE}$ the negative pressure that makes the tiny voids due to thermal 
fluctuations grow.  In the dark energy domination epoch, this pressure is given by $p_{\rm DE} = w
\rho_{\rm DE}$, where $w$ is the effective equation of state and $\rho_{\rm DE} \sim 10^{-12}$ eV$^{4}$ 
is the present dark energy density.  A criterion for the absence of cavitation can be phrased in the 
following way: there is no cavitation in the system if the tensile strength is larger than the maximal 
outward pressure:
\begin{eqnarray}
\label{eq:Criterion_cavitation}
p_{b\;\rm max} & > & |w|\rho_{\rm DE}.
\end{eqnarray}
This condition must be satisfied for a sufficiently long time in order to have dark energy domination 
without the fluid to break apart.  The tensile strength $p_{b\;\rm max}$ is a property of the fluid and may 
depend on $\lambda$, $m$ and the dimensionless inverse temperature $x = m_{0}/T_{s}$.  Thus the 
above criterion for no cavitation is a constraint on the parameters of the model.  

We can roughly estimate the scalar fluid tensile strength using the theory of homogeneous nucleation 
(e.g.~\cite{Brennen_1995}).  
The result is that the tensile strength is given by:
\begin{eqnarray}
\label{eq:Tensile_strength_method1_1}
p_{b\;\rm max} & = & \left(\frac{16\pi S^{3}}{3T_{s}\ln(J_{0}/J)} \right)^{1/2},
\end{eqnarray}
where $S$ is the surface tension and $J$ is the nucleation rate.  Since the above relation depends 
logarithmically on the nucleation rate, we expect a very weak dependence on $J$.  Thus for estimate 
purposes, we set the logarithm to one.  The surface tension is a force per unit length; we thus expect it to 
be proportional to the coupling $\lambda$ and a quantity that has dimensions of [Energy]$^{3}$.  The 
two relevant energy scales in the problem are $m_{0}$ and $T_{s}$.  From dimensional analysis, we 
obtain four possible forms for the surface tension: $S\sim \lambda T_{s}^{3}$, $S\sim \lambda 
m_0 T_{s}^{2}$, $S\sim \lambda m_0^{2}T_{s}$ and $S\sim \lambda m_0^{3}$.  The first one is expected to be 
valid only at ultra-relativistic energies and we discard it.  The last one is unrealistic because of its lack of 
temperature dependence.  Empirical laws show that the typical behavior of surface tension is linear in 
temperature.  We thus use the form $S\sim \lambda m_0^{2}T_{s}$ in the following analysis.  Plugging $S\sim \lambda m_0^{2}T_{s}$ in Eq. (\ref{eq:Tensile_strength_method1_1}) we get:
\begin{eqnarray}
\label{eq:Tensile_strength_method1_2}
p_{b\;\rm max} & \sim & \left(\frac{16\pi}{3}\right)^{1/2}\frac{\lambda^{3/2}m_0^{4}}{x}.
\end{eqnarray}
Combining this with the no cavitation criterion, we obtain the following constraint on the model 
parameters:
\begin{eqnarray}
\label{eq:Criterion_cavitation_method1}
\frac{\lambda^{3/2}m_0^{4}}{x} & > & \left(\frac{3}{16\pi}\right)^{1/2}|w|\rho_{\rm DE}.
\end{eqnarray}
%
%
Cavitation is also studied in the context of relativistic heavy ion collisions~
\cite{Rajagopal_Tripuranemi_2009}.  In this study, it is stated that any hydrodynamic simulations should 
be stopped when the pressure goes below zero because of cavitation.  We argue here that, just like in 
other engineering systems, a fluid can sustain some negative pressure without cavitating.

\end{description}

\section{Background evolution}
\label{sec:Background_evolution}

\subsection{Implementation in \texorpdfstring{\tt CLASS}{CLASS}}
\label{sec:Implementation_CLASS}

The pressure, energy density and bulk viscosity are all functions of the dimensionless inverse 
temperature $x = m_{0}/T_s$ (c.f. eqs.~\ref{eq:Pressure_scalar}--\ref{eq:Bulk_viscosity_scalar}):
\begin{eqnarray}
\label{eq:Pressure_scalar_x}
p_{s}(x) & = & \frac{m_{0}^{4}}{2\pi^{2}}\frac{K_{2}(x)}{x^{2}}, \\
\label{eq:Energy_density_scalar_x}
\rho_{s}(x) & = & \frac{m_{0}^{4}}{2\pi^{2}}\left(\frac{K_{2}(x)}{x^{2}} - \frac{K_{2}'(x)}{x}\right), \\
\label{eq:Bulk_viscosity_scalar_x}
\zeta & = & \left(\frac{\tilde{m}^{4}}{\lambda^{4}m_{\rm th}}\right) \left(x\frac{m_{\rm th}}{m_{0}}
\right)^{\kappa_{3}} e^{\kappa_{1}} e^{\kappa_{2} (m_{\rm th}/m_{0})x},
\end{eqnarray}
where primes denote derivatives with respect to $x$ and the subscript $s$ relates to the scalar particles.  

The energy conservation equation relates $\dot{\rho}_s$ to $\rho_s$ and $p_{\rm eff}=p_s-3H\zeta$. The ususal 
procedure is to use the equation of state to express the pressure in terms of $\rho$ and then solve for $\rho$.  In our case, the results for the pressure, energy density and bulk viscosity are all expressed in terms of $x$.  To integrate the equations, we re-write the LHS of the energy conservation equation~
\ref{eq:Energy_conservation_perfect} as $(d\rho_{s}/dx)(dx/dt)$, where the derivative $d\rho_{s}/dx$ is 
known.  The resulting background equations requiring a numerical integration over time are:
\begin{eqnarray}
\label{eq:Background_equation_rho}
\dot{a} & = & a H~,\\
\label{eq:Background_equation_x}
\dot{x} & = & -3H\left(\frac{\rho_{s} + p_{s} - 3H\zeta}{\rho_{s}'}\right)~,
\end{eqnarray}
where $H$ is inferred from the Friedmann equation, including a contribution $\rho_s$ to the total energy 
density, and $\rho_{s}'$ is given by:
\begin{eqnarray}
\label{eq:Derivative_energy_density}
\rho_{s}'(x) & = & \frac{m_{0}^{4}}{2\pi^{2}}\left(-\frac{2K_{2}(x)}{x^{3}} + \frac{2K_{2}'(x)}{x^{2}} - 
\frac{K_{2}''(x)}{x}\right).
\end{eqnarray}
Equations~\ref{eq:Background_equation_rho} and~\ref{eq:Background_equation_x} are integrated 
using the Cosmic Linear Anisotropy Solving System ({\tt CLASS}) code~\cite{Lesgourgues_2011,Blas_etal_2011} modified to take into account the new scalar energy 
component.  

The minimal extension of the flat $\Lambda$CDM model in which the cosmological constant is replaced 
by a viscous scalar fluid has eight free parameters, namely the three fundamental parameters of the 
scalar sector ($m_{0}$, $\lambda$, $\epsilon$) and five parameters describing the other sectors.  These five parameters are chosen to be ($\Omega_{b}$, $h$, $A_s$, $n_s$, $\tau$) (baryon density fraction, 
reduced Hubble parameter, primordial spectrum amplitude and tilt, reionization optical depth). In principle, 
each combination of ($m_{0}$, $\lambda$, $\epsilon$, $h$) leads to a unique value of the current 
scalar density fraction $\Omega_s$, from which one can infer the CDM density fraction $\Omega_{cdm} = 1 - 
\Omega_{b}-\Omega_{s}$. The fact that the universe is flat imposes a bound in the 
($m_{0}$, $\lambda$, $\epsilon$, $h$)  space ensuring that $\Omega_s\leq1$. 

In order to compare our model with observations, it is much more 
convenient to use $\Omega_s$ or $\Omega_m=1-\Omega_s$ as one of the model parameters.
In that case, $m_0$ can be seen as a function of ($\Omega_m$, $\lambda$, $\epsilon$, $h$).
This parametrisation is the one that we implement in {\tt CLASS}. For each model parameters
\begin{equation}
(\Omega_{m},~\Omega_{b},~h,~A_s,~n_s,~\tau,~\lambda,~\epsilon)~,
\label{eq:parameters}
\end{equation}
the code searches for the value of $m_0$ leading to the correct relic density with a simple bisection algorithm.
This involves running the {\tt CLASS} module {\tt background.c} several times with different $m_0$'s until the correct mass is found up to a given accuracy.  The other modules are run after the obtention of the correct mass.

\subsection{Results for the background evolution}
\label{sec:Results_background}

Figure~\ref{fig:background_evolution} shows the evolution of the densities, temperature and effective pressure in the
minimal model including a viscous fluid described by the eight parameters listed above (note that only five of them are relevant for the background evolution). Solid lines correspond to the choice
\begin{equation}
(\Omega_{\rm m},~\Omega_{\rm b},~h,~\lambda,~\epsilon)
= (0.28, 0.05~, 0.72, 0.25, 0.7),
\end{equation}
giving $m_0=1.0$~eV with a viscous fluid temperature $T_{s}=0.7\,T_\gamma$ of the same order as the neutrino temperature as long as the fluid behaves like radiation. Dashed lines correspond to the same parameters as above except for $\epsilon=0.1$, implying that a number $g_* \sim 100$ of relativistic degrees of freedom annihilate after the scalar-photon decoupling, and transfer their entropy to the photons and particles in thermal equilibrium with them.  Finally, dotted lines correspond to a model with the same parameters as above except for $\lambda=10^{-4}$, giving $m_0 = 2.7\times10^{-2}$~eV.

On the left panel, we see that scalar species are initially relativistic with $\rho_{\rm s} \propto a^{-4}$ and $T_{s} \propto a^{-1}$.
This remains true as long as $x\ll1$ ($T_{s} \gg m_0$) and bulk viscosity is negligible.  The contribution of the scalar fluid to the commonly used ``effective neutrino number'' $N_{\rm eff}$ is given by
\begin{equation}
\Delta N_{\rm eff} = \frac{1}{2} \times \frac{8}{7} \left(\frac {11}{4}\right)^{4/3} \epsilon^4 = 2.2 \epsilon^4,
\end{equation}
where the factor is obtained from the neutrino-to-photon temperature ratio assumed in the usual definition of $N_{\rm eff}$ and from the fact that we are considering bosons ($\Delta N_{\rm eff}=1$ corresponds to the case of one fermion plus its anti-particle).

When $x \sim 1$ ($T_{s} \sim m_0$), the bosons become non-relativistic and their density starts to dilute like that of ordinary matter ($\rho_{\rm s} \propto a^{-3}$). This is clearly visible for the model with solid or dashed lines in figure~\ref{fig:background_evolution}. In the case of collisionless species like massive neutrinos, the temperature keeps decreasing like $a^{-1}$ during the non-relativistic regime.  This is not true for self-interacting species like the scalar fluid considered here.
An analytic study of the evolution equations in the limit $x\gg1$, $\zeta=0$ shows that the temperature evolves according to $T_{s}^{3/2} e^{-m_0/T_{s}}\propto a^{-3}$ during this stage.

We also see that the viscosity $\zeta$ increases with time. At some point, the terms $3 H \zeta$ and $p_s$ become of the same order. After that moment, the effective pressure $p_{\rm eff}=p_s-3H\zeta$ changes sign and finally approaches $-\rho_s$.  The system
then tends towards a fixed point with constant values of $\rho_s$, $p_s$, $H$, $\zeta$ and $T_s$ such that $p_{\rm eff}=-\rho_s$ (and $p_s \ll 3 H \zeta$).
As expected, this mechanism mimics dark energy at least at the level of the background evolution.

\subsection{Parameter dependence}
\label{sec:Parameter_dependence}

\begin{figure}
\centering
\begin{tabular}{c}
\includegraphics[width=0.5\textwidth]{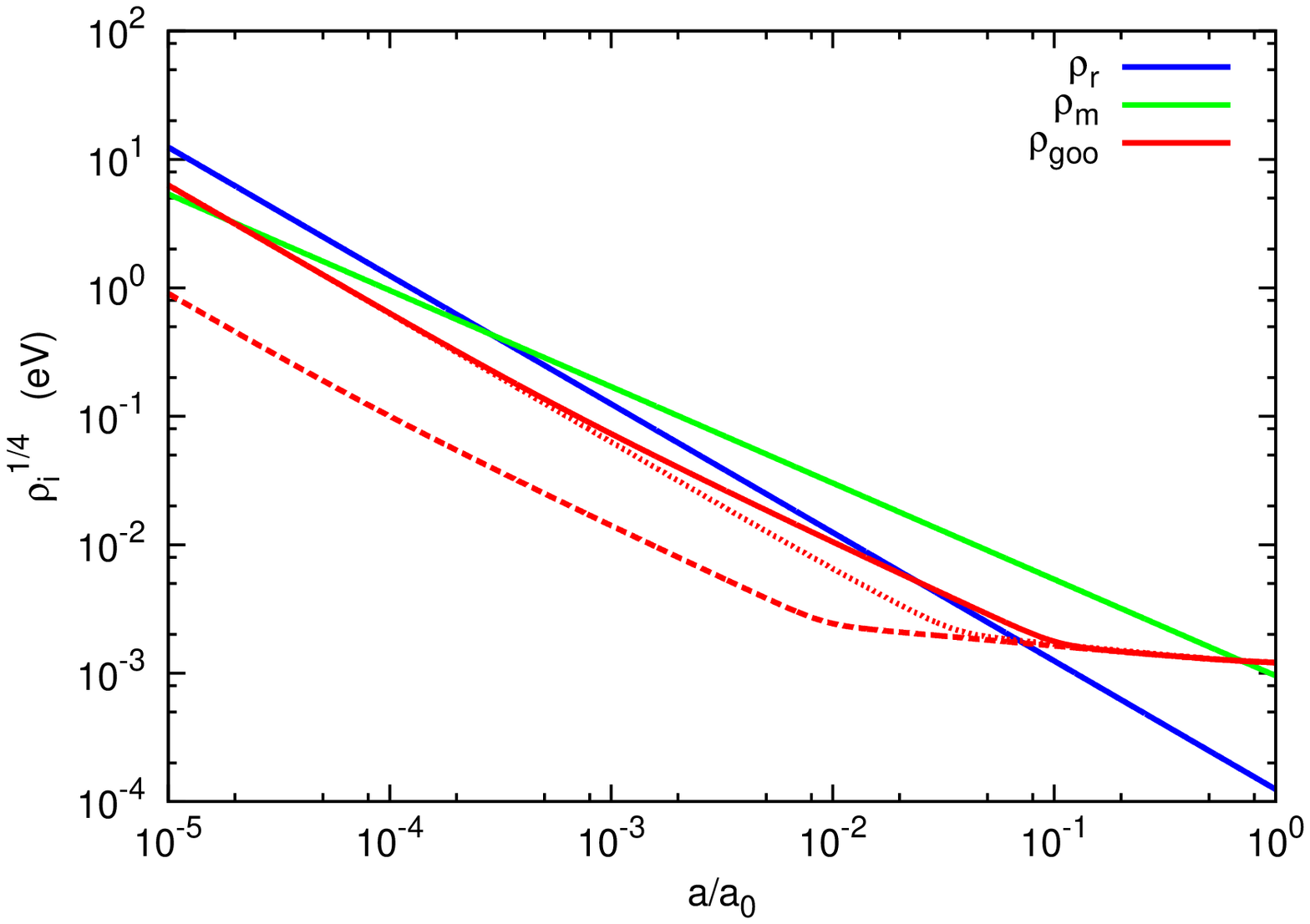}
\includegraphics[width=0.5\textwidth]{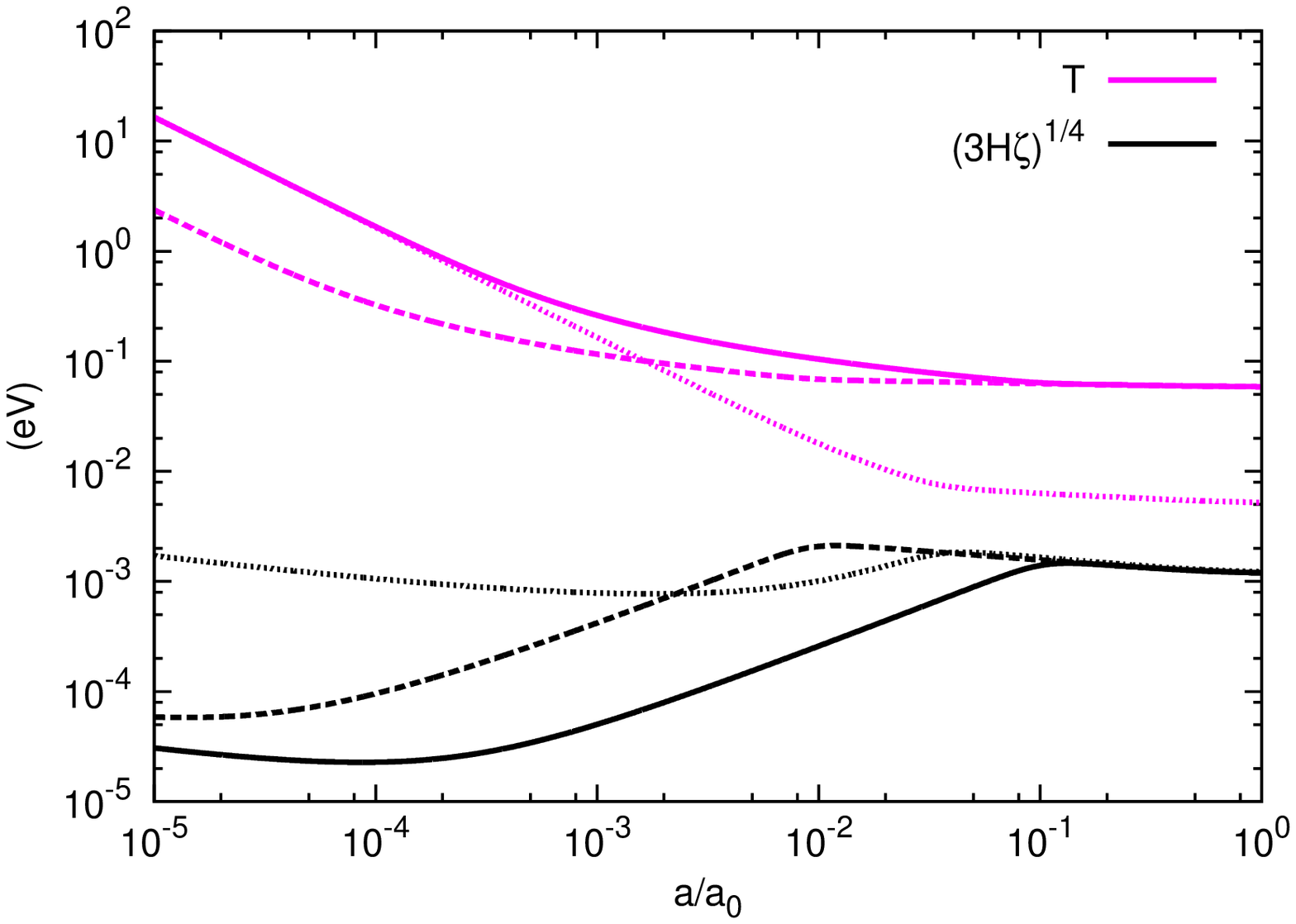}
\end{tabular}
\caption{Left: Evolution of the radiation, matter and viscous fluid densities in a model with $h=0.72$ and $\Omega_m=0.28$.  Solid lines correspond to the parameter choice $(\epsilon, \lambda)=(0.7,0.25)$ leading to $m_0=1.0$~eV.  The dotted (resp. dashed) line is a variant with $(\epsilon, \lambda)=(0.1,0.25)$ (resp. $(0.7, 10^{-4}$)) leading to $m_0=1.0$~eV (resp. $m_0=2.7\times 10^{-2}$~eV).  Right: For the same models, evolution of the viscous fluid temperature and of the correction to the effective pressure induced by viscosity.}
\label{fig:background_evolution}
\end{figure}

It is interesting to compare the three models displayed in figure~\ref{fig:background_evolution}.
In the model with a smaller $\epsilon$ (dashed lines), the fluid density and temperature are initially smaller
(in this case, $N_{\rm eff}\simeq3.04$ like in the minimal $\Lambda$CDM scenario). Since the mass
is unchanged with respect to the reference model with solid lines,
the transition to the non-relativistic regime takes place at the same temperature $T_s\sim1$~eV, corresponding to an earlier time. As a consequence of the smaller temperature, viscosity becomes important earlier and the scalar fluid starts its dark energy behavior earlier.
In the model with a smaller value of $\lambda$ (dotted lines), and also a smaller value of $m_0$ corresponding to the same relic density $\Omega_s$, the early and late behaviors are similar to those of the reference model. However, we do not observe an intermediate non-relativistic stage with $\rho_s \propto a^{-3}$, because the time at which $T_{s} \sim m_0$ coincides with that at which viscosity becomes important: the fluid goes directly from a radiation-like to a dark-energy-like behavior.

The relation between $\lambda$ and $m_0$ for a given ($\Omega_s$, $h$) or ($\Omega_m$, $h$) is difficult to estimate analytically.  In principle, this relation can be obtained by noticing that close to the fixed point one has:
\begin{equation}
\rho_s^0 = \Omega_s \rho_c^0 \simeq -p_{\rm eff}~.
\end{equation}
The effective pressure is then dominated by the bulk velocity contribution, so we can rewrite these two
equalities (where the second one is approximate) as:
\begin{equation}
\rho_s^0 = \Omega_s \frac{3 H_0^2}{8 \pi G} \simeq 3 H_0 \zeta~.
\end{equation}
Replacing $\rho_s$ and $\zeta$ by their expression in terms of $x$ and other fundamental parameters (c.f. eqs. \ref{eq:Energy_density_scalar_x}, \ref{eq:Bulk_viscosity_scalar_x}), we obtain two independent relations between $(x, m_0, \lambda, H_0, \Omega_s)$. These two equations can be combined in order to extract $x$ and find the temperature at the fixed point as a function of $(m_0, \lambda, H_0, \Omega_s)$.
Finally, $x$ can be replaced in any of the two equations in order to obtain a relation between
$(m_0, \lambda, H_0, \Omega_s)$.  Note that $\epsilon$ plays no role in this calculation. 

\begin{figure}
\centering
\begin{tabular}{c}
\includegraphics[width=0.5\textwidth]{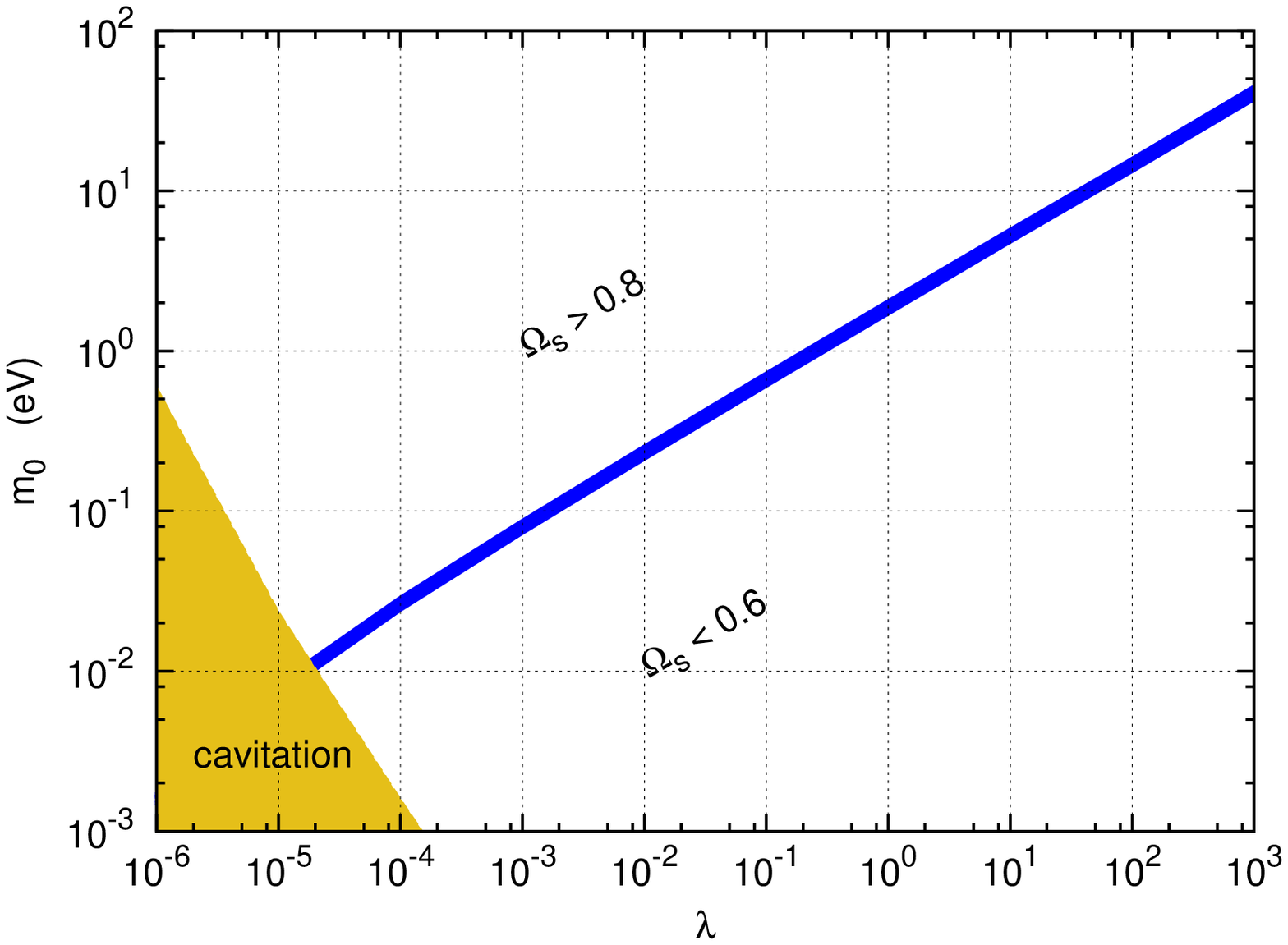}
\includegraphics[width=0.5\textwidth]{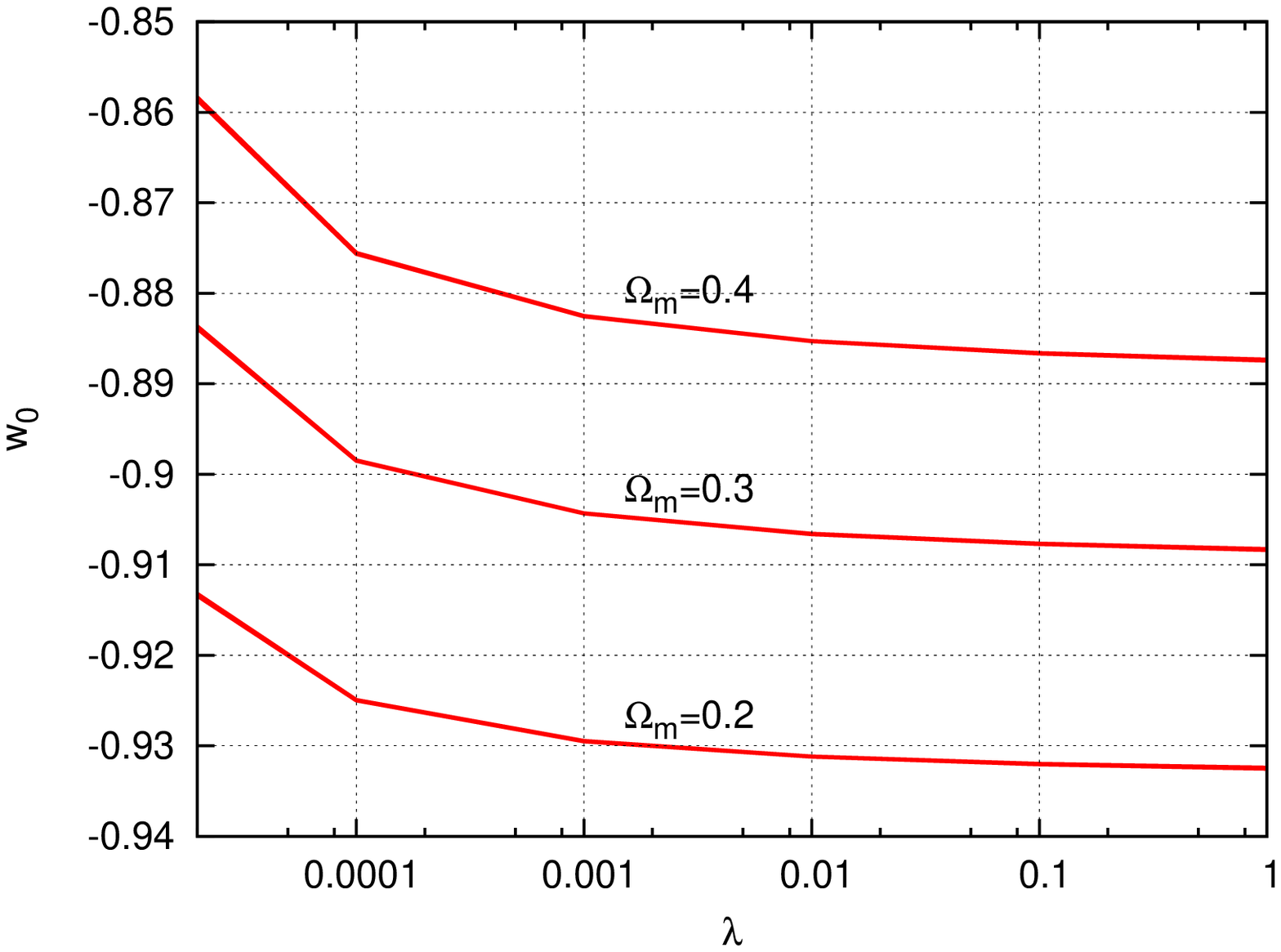}
\end{tabular}
\caption{Left: Relation between $m_{0}$ and $\lambda$ in a flat universe with $0.2<\Omega_m<0.4$ (i.e. with $0.6<\Omega_\goo<0.8$), assuming $0.01<\epsilon<1$ and $0.65<h<0.75$. The shaded region is excluded by constraints from hydrodynamics (cavitation). Right: For the same range of parameters $\epsilon$ and $h$, current value of the equation of state parameter of the viscous fluid as a function of $\lambda$, for three different values of $\Omega_m$.}
\label{fig:Relation_between_m_lambda}
\end{figure}

This system of equation is difficult to solve analytically.  However we obtain such a relation numerically, thanks to the bisection algorithm implemented in {\tt CLASS}. Namely, we vary $\lambda$ in the range $[10^{-10},10^3]$, $h$ in the range $[0.65, 0.75]$, $\Omega_m=1-\Omega_s$ in the range
$[0.2,0.4]$ and obtain the corresponding $m_0$ in each case.
We show in the left panel of figure~\ref{fig:Relation_between_m_lambda} the corresponding
region in ($\lambda$, $m_0$) space  (thin blue band). This region
is compatible with hydrodynamical constraints only in the range
$\lambda > 2\times 10^{-5}$ (otherwise the fluid is broken apart by cavitation),
corresponding to $m_0 > 0.01$~eV.  The fluid is always in local equilibrium in this range of coupling.  The scaling between $\lambda$ and $m_0$ changes around the value $\lambda \sim 10^{-5}$, but in the physically interesting range $\lambda \gg 10^{-5}$ we find a fitting formula accurate at the percent level:
\begin{equation}
m_0=1.88 ~\lambda^{0.450} \left(\frac{\Omega_s}{0.72}\right)^{0.295} 
\left(\frac{h}{0.72}\right)^{0.445} {\rm eV}~.
\end{equation}
Note that the validity of the model for a large self-coupling constant is far from obvious, since the expression
for bulk viscosity is obtained using perturbation theory and assumes $\lambda<1$.
If we impose a theoretical prior $\lambda \leq 1$, the allowed range for the mass reduces to
$m_0 \in [0.01, 2]$~eV.

We checked numerically that the above relation between $m_0$ and $\lambda$ is not affected by
$\epsilon$, as expected from our previous reasoning.  Moreover, the current ratio $x^0=m_0/T_s^0$ depends on $(\lambda, H_0, \Omega_s)$ but not on $\epsilon$. Hence the $\epsilon$ parameter is only important at early time through its impact on the total radiation density.

We see from figure~\ref{fig:background_evolution} that the viscous fluid does not reach the fixed point (or stabilize with a constant energy density) immediately.  This suggests that our model predicts a value of the equation-of-state parameter $w=p_s/\rho_s$ larger than -1 today and slightly varying with time.
A priori, the current value $w_0$ could depend on $(\lambda, H_0, \Omega_s)$ and even slightly on $\epsilon$.
A careful numerical investigation illustrated by the right panel of figure~\ref{fig:Relation_between_m_lambda} shows that $w_0$ 
depends only on $\lambda$ and $\Omega_s$ (or $\Omega_m$).  The same appears to be true for its time-derivative. This seems to imply that the present model is an example of a bulk viscosity-induced dark energy model that does not ``cross the phantom divide''~\cite{Nojiri_Odintsov_2005,Capozziello_etal_2005,Brevik_Gorbunova_2005}.

In summary of this subsection, we find that the model under consideration has the same 
background evolution as a minimal $\Lambda$CDM model extended in the following way:
\begin{itemize}
	\item It has extra relativistic degrees of freedom at early times, controlled by the parameter $\epsilon$.  The equivalent effective neutrino number is
\begin{equation}
N_{\rm eff}= 3.04 + 2.2 \epsilon^4~.
\label{eq:fit_neff}
\end{equation}
If scalar particles decouple from the rest of the plasma very late (but before electron-positron annihilation), $\epsilon$ may be in the range $[0.5, 0.7]$ and produce a significant increase of $N_{\rm eff}$ \footnote{Indeed, if the bosons decouple before electron-positron annihilation, their temperature is at most equal to that of neutrinos, $T_{\nu} = (4/11)^{1/3} T_{\gamma} \simeq 0.7 T_{\gamma}$. In principle, it would be possible to assume that our bosons decouple from the plasma after electron-positron annihilation, corresponding to $\epsilon$ in the range [0.7,1].  The physics of Big Bang Nucleosynthesis and photon recombination would then be affected in a more non-trivial way than through $N_{\rm eff}$.  We do not consider this case for simplicity.}. If they decouple early enough such that $\epsilon < 0.5$, the increase in $N_{\rm eff}$ is negligible given the accuracy of future CMB and LSS experiments.
	\item It incorporates slowly decaying dark energy, so our model mimics the so-called wCDM model~\cite{Chevallier_Polarski_2000,Linder_2002} rather than $\Lambda$CDM. The value of $w$ today is found to be controlled by ($\Omega_s$, $\lambda$) or equivalently by ($\Omega_m$, $\lambda$), and the following fit is accurate to better than one percent:
\begin{equation}
w_0=-0.9085+0.21(\Omega_m-0.3)+3 \lambda^{-0.4} 10^{-4}~.
\label{eq:fit_w0}
\end{equation}

	\item For a very precise description, one should take into account the fact that
the parameter $w$ varies slowly with time during dark energy domination. If we try to capture 
this small redshift variation with a standard first-order expansion of the type $w(a)=w_0+w_a(1-a/a_0)$ \cite{Chevallier_Polarski_2000,Linder_2002}, we find numerically that the fit
\begin{equation}
w_a=0.129+0.02(\Omega_m-0.3)+\lambda^{-0.5} 10^{-4},
\label{eq:fit_wa}
\end{equation}
is accurate up to 5\%.

\end{itemize}
It remains to be checked that the equivalence between our model and such an extended wCDM model
is valid also at the level of perturbations. The study of linear perturbations is the object of section~\ref{sec:Linear_perturbations}.

Since for $\lambda \gg 10^{-4} $ the scalar fluid undergoes a non-relativistic regime with $\rho_s \propto a^{-3}$ prior to the dark energy regime, one could think that it could play alternatively the role of dark matter and dark energy. At the level of the background evolution, this could be achieved by setting $\Omega_{cdm}=0$, $\Omega_m=\Omega_s+\Omega_{b}=1$, and by increasing $\lambda$ (and hence $m_0$) until the scalar density during the non-relativistic regime matches the usual CDM density. We find that this is indeed possible, but for values of $\lambda$ much larger than one.  We conclude that our bulk viscosity model cannot explain dark matter and dark energy at the same time, but in principle one could try to extend this model with such a goal.

\section{Linear perturbations}
\label{sec:Linear_perturbations}

As pointed out in section~\ref{sec:Validity_hydrodynamics}, it is important to study linear perturbations 
around the background solution to discuss the issue of small inhomogeneities and instabilities.  In the 
following we derive the equations for linear perturbations (including viscous effects) in the synchronous 
gauge, since it is the gauge that is normally used in numerical simulations.

\subsection{Equations for the perturbations}
\label{sec:Equations_perturbations}

Our starting point are the standard evolution equations for scalar perturbations in the synchronous 
gauge (we follow the conventions of Weinberg~\cite{Weinberg_2008}):
\begin{eqnarray}
\label{eq:Linear_perturbations_rho_1}
\delta\dot{\rho} + 3\frac{\dot{a}}{a}(\delta\rho + \delta p) -\frac{k^{2}}{a^{2}}(\rho + p)\delta U + (\rho + p)
\psi - k^{2}\frac{\dot{a}}{a}\pi^{S} & = & 0, \\
\label{eq:Linear_perturbations_U_1}
(\rho + p)\delta\dot{U} + \left((\dot{\rho} + \dot{p}) + 3\frac{\dot{a}}{a}(\rho + p)\right)\delta U + \delta p - 
k^{2}\pi^{S} & = & 0, \\
\label{eq:Linear_perturbations_psi_1}
\dot{\psi} + 2\frac{\dot{a}}{a}\psi + 4\pi G(\delta\rho + 3\delta p - k^{2}\pi^{S}) & = & 0,
\end{eqnarray}
where $\rho$ and $p$ are the background values for the density and pressure, $\delta\rho$ and $\delta 
p$ are the pertubations in density and pressure, $\delta U_{i} = \partial_{i}\delta U + \delta U_{i}^{V}$ are 
the velocity perturbations (where $\partial_{i}\delta U_{i}^{V} = 0$), $\psi$ is a linear combination of metric 
perturbations, $\pi^{S}$ is the anisotropic stress tensor, $k$ is a wavevector and dots represent 
derivatives with respect to proper time.  Equations~\ref{eq:Linear_perturbations_rho_1} and~
\ref{eq:Linear_perturbations_U_1} correspond to energy conservation and momentum conservation, 
respectively.  There is one set of energy-momentum conservation equations for each type of matter that 
does not exchange energy and momentum with other constituents.  In the following we present the 
treatment of linear perturbations for the added scalar matter, but it must be borne in mind that we 
consider all linear perturbations in our final analysis.

By assumption the scalar fluid in our model only interacts gravitationally with other particles.  Thus we 
can write a set of energy-momentum conservation equations for the scalar perturbations $\delta \rho_{s}
$ and $\delta U_{s}$.  Since equations~\ref{eq:Linear_perturbations_rho_1}--
\ref{eq:Linear_perturbations_psi_1} do not depend on the particular form of the energy-momentum 
tensor, we make the replacements $p \rightarrow p_{\rm eff}$ and $\delta p \rightarrow \delta p_{\rm eff}$ 
to account for bulk viscous effects.  We also assume that the anisotropic stress tensor $\pi^{S}$ for our 
scalar fluid is zero.  There is no data to constrain this number and we make this assumption to simplify 
the equations.  The evolution equations of perturbations for our scalar fluid read:
\begin{eqnarray}
\label{eq:Linear_perturbations_rho_2}
\delta\dot{\rho}_{s} + 3\frac{\dot{a}}{a}(\delta\rho_{s} + \delta p_{\rm eff}) -\frac{k^{2}}{a^{2}}(\rho_{s} + 
p_{\rm eff})\delta U_{s} + (\rho_{s} + p_{\rm eff})\psi & = & 0, \\
\label{eq:Linear_perturbations_U_2}
(\rho_{s} + p_{\rm eff})\delta\dot{U}_{s} + \left((\dot{\rho}_{s} + \dot{p}_{\rm eff}) + 3\frac{\dot{a}}{a}(\rho_{s} 
+ p_{\rm eff})\right)\delta U_{s} + \delta p_{\rm eff} & = & 0, \\
\label{eq:Linear_perturbations_psi_2}
\dot{\psi} + 2\frac{\dot{a}}{a}\psi + 4\pi G(\delta\rho_{s} + \delta\rho_{\rm other} + 3\delta p_{\rm eff} + 
3\delta p_{\rm other}) & = & 0,
\end{eqnarray}
where the subscript ``$s$'' refer to the scalar fluid and ``other'' to the other fluids present (photons, neutrinos, baryons, 
dark matter).  We note that all fluid perturbations are necessary to obtain the correct metric perturbations.

In our model, the forms of $\rho_{s}$ and $p_{\rm eff}$  are given in eqs.~\ref{eq:Pressure_scalar_x}--
\ref{eq:Bulk_viscosity_scalar_x}.  We note that the energy density is only a function of $x$. So we can 
express a perturbation in energy density solely as a perturbation in $x$ (or equivalently a change in 
temperature):
\begin{eqnarray}
\label{eq:Perturbed_density}
\delta\rho_{s} & = & D(x) \, \delta x,
\end{eqnarray}
where the function $D(x)=\rho_s'(x)$ is given by eq.~\ref{eq:Derivative_energy_density}.  The change in 
effective pressure is a bit more tricky, since $p_{\rm eff}$ also depends on $D_{\gamma}U^{\gamma}$.  
We write:
\begin{eqnarray}
\label{eq:Perturbed_effective_pressure}
\delta p_{\rm eff} & = & \delta\left(p_{s} - \zeta D_{\gamma}U^{\gamma}\right) \nonumber \\
									 & = & \delta p_{s} - \delta\zeta (D_{\gamma}U^{\gamma}) 
- \zeta \, \delta (D_{\gamma}U^{\gamma}) \nonumber \\
									 & = & \left(\frac{\partial p_{s}}{\partial x}\right)\delta x - 
3\frac{\dot{a}}{a}\left(\frac{\partial\zeta}{\partial x}\right)\delta x - \zeta \, \delta (D_{\gamma}U^{\gamma}) 
\nonumber \\
									 & \equiv & P_{\rm eff}(x)\delta x - \zeta \, \delta 
(D_{\gamma}U^{\gamma}),
\end{eqnarray}
where the function $P_{\rm eff}(x)$ can be obtained from eqs.~\ref{eq:Pressure_scalar_x} and 
\ref{eq:Bulk_viscosity_scalar_x}.  To obtain $\delta (D_{\gamma}U^{\gamma})$, we write:
\begin{eqnarray}
\delta (D_{\nu}U^{\nu}) & = & \delta\left(\frac{\partial U^{\nu}}{\partial x^{\nu}} + \Gamma_{\nu
\lambda}^{\nu}U^{\lambda}\right) \nonumber \\
											  & = & \frac{\partial (\delta U^{\nu})}{\partial 
x^{\nu}} + \Gamma_{\nu\lambda}^{\nu}(\delta U^{\lambda}) + (\delta\Gamma_{\nu
\lambda}^{\nu})U^{\lambda} \nonumber \\
											  & = & \frac{\partial (\delta U^{\nu})}{\partial 
x^{\nu}} + 3\frac{\dot{a}}{a}(\delta U^{0}) + (\delta\Gamma_{\nu\lambda}^{\nu})U^{\lambda},
\end{eqnarray}
where in the last line we used $\Gamma_{\nu\lambda}^{\nu}(\delta U^{\lambda}) = 
(\Gamma_{0\lambda}^{0} + \Gamma_{i\lambda}^{i})\delta U^{\lambda} = \Gamma_{i0}^{i}\delta U^{0} + 
\Gamma_{ij}^{i}\delta U^{j} = (3\dot{a}/a)\delta U^{0}$ when spatial curvature is zero. The 4-velocity is 
$U^{\mu} = (-1,v^{i})$ and thus $\delta U^{0} = 0$.  Keeping only linear terms in velocity and 
perturbations and using the perturbed Christoffel symbols in the synchronous gauge~\cite{Weinberg_2008}, we obtain:
\begin{eqnarray}
\delta (D_{\nu}U^{\nu}) & = & a^{-2}\nabla^{2}\delta U + \psi.
\end{eqnarray}
The perturbation of the effective pressure can thus be expressed in terms of temperature, velocity and 
metric perturbations:
\begin{eqnarray}
\label{eq:Perturbation_effective_pressure}
\delta p_{\rm eff} & = & P_{\rm eff}(x)\delta x - \zeta(a^{-2}\nabla^{2}\delta U + \psi).
\end{eqnarray}
Using eqs.~\ref{eq:Perturbed_density} and~\ref{eq:Perturbation_effective_pressure}, we can re-write the 
evolution equations for the perturbations \ref{eq:Linear_perturbations_rho_2}--
\ref{eq:Linear_perturbations_psi_2} as:
\begin{equation}
\label{eq:Linear_perturbations_rho_3}
\delta\dot{x} + \left[\frac{\dot{D} + 3H(D+P_{\rm eff})}{D}\right]\delta x + \left[\frac{3\zeta H\frac{k^{2}}{a^{2}} - 
\;\;\frac{k^{2}}{a^{2}}(\rho_{s} + p_{\rm eff})}{D}\right]\delta U_{s} + \left[\frac{(\rho_{s} + p_{\rm eff} - 3H\zeta)}
{D}\right]\psi = 0, 
\end{equation}
\begin{equation}
\label{eq:Linear_perturbations_U_3}
\hspace{1.08in} (\rho_{s} + p_{\rm eff})\delta\dot{U}_{s} + P_{\rm eff}\delta x + \left[(\dot{\rho}_{s} + \dot{p}_{\rm eff}) + 
3H(\rho_{s} + p_{\rm eff}) + \zeta \frac{k^{2}}{a^{2}}\right]\delta U_{s} - \zeta\psi = 0, \\
\end{equation}
\begin{equation}
\label{eq:Linear_perturbations_psi_3}
\dot{\psi} + \left[4\pi G(D + 3P_{\rm eff})\right]\delta x + \left[12\pi G\zeta \frac{k^{2}}{a^{2}}\right]\delta U_{s} + \left[2(H 
- 6\pi G\zeta)\right]\psi + 4\pi G (\delta\rho_{\rm other} + 3\delta p_{\rm other}) = 0.
\end{equation}
%
%
These are the evolution equations for scalar perturbations for the viscous (scalar) fluid.  We integrate 
them using {\tt CLASS} and the results are shown in section~\ref{sec:Results_perturbations}.  For completness and in order to make our results easily reproducible, we provide in Appendix \ref{sec:appendix} a complete list of the equations added to the {\tt CLASS} perturbation module.  Essentially, these are
 equations~\ref{eq:Linear_perturbations_rho_2}, \ref{eq:Linear_perturbations_U_2} and \ref{eq:Perturbation_effective_pressure} translated into the notations of Ma \& Bertschinger~\cite{Ma:1995ey}, which are used throughout the {\tt CLASS} code.

\subsection{Results for the perturbation evolution}
\label{sec:Results_perturbations}

\begin{figure}
\centering
\begin{tabular}{c}
\includegraphics[width=0.5\textwidth]{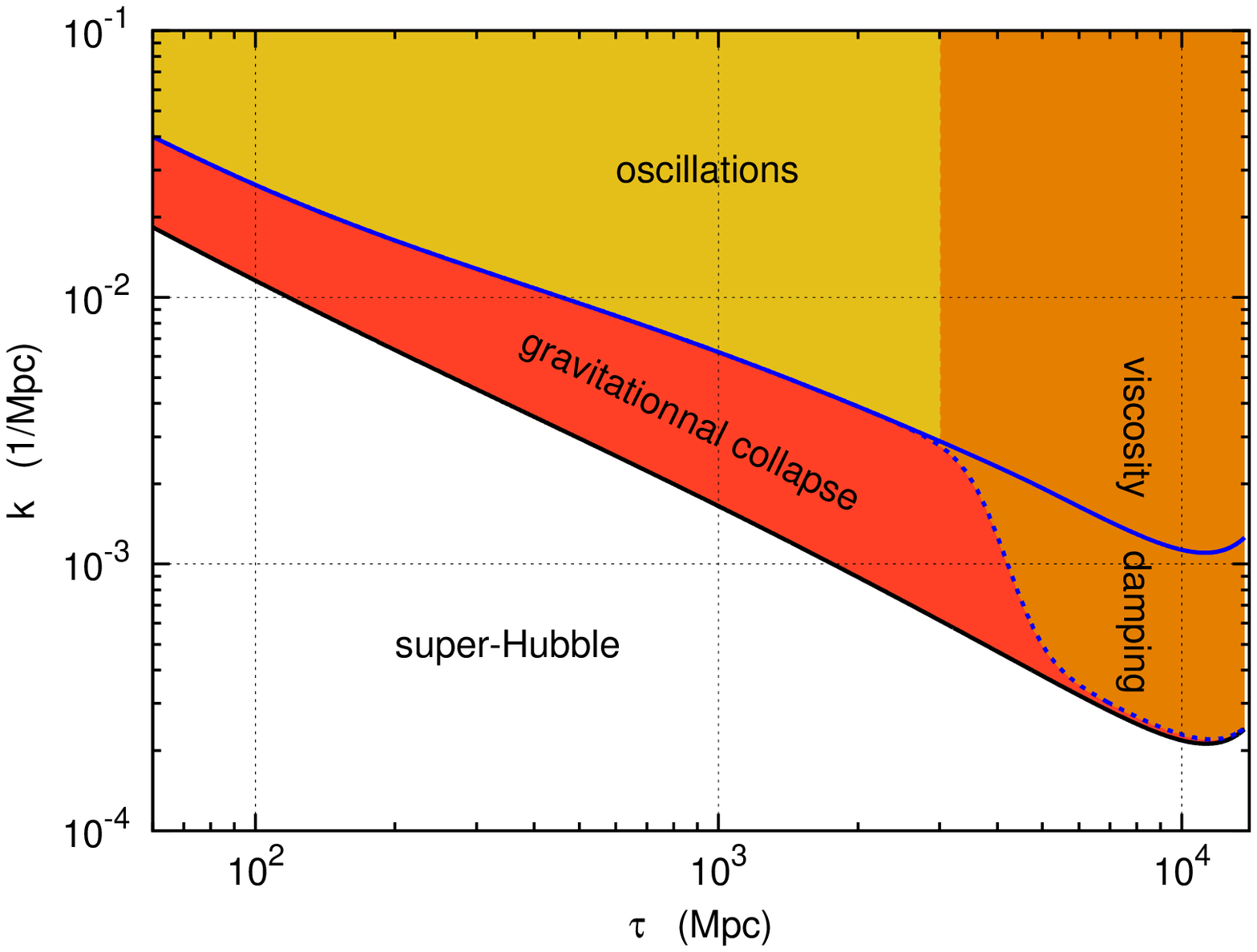}
\includegraphics[width=0.5\textwidth]{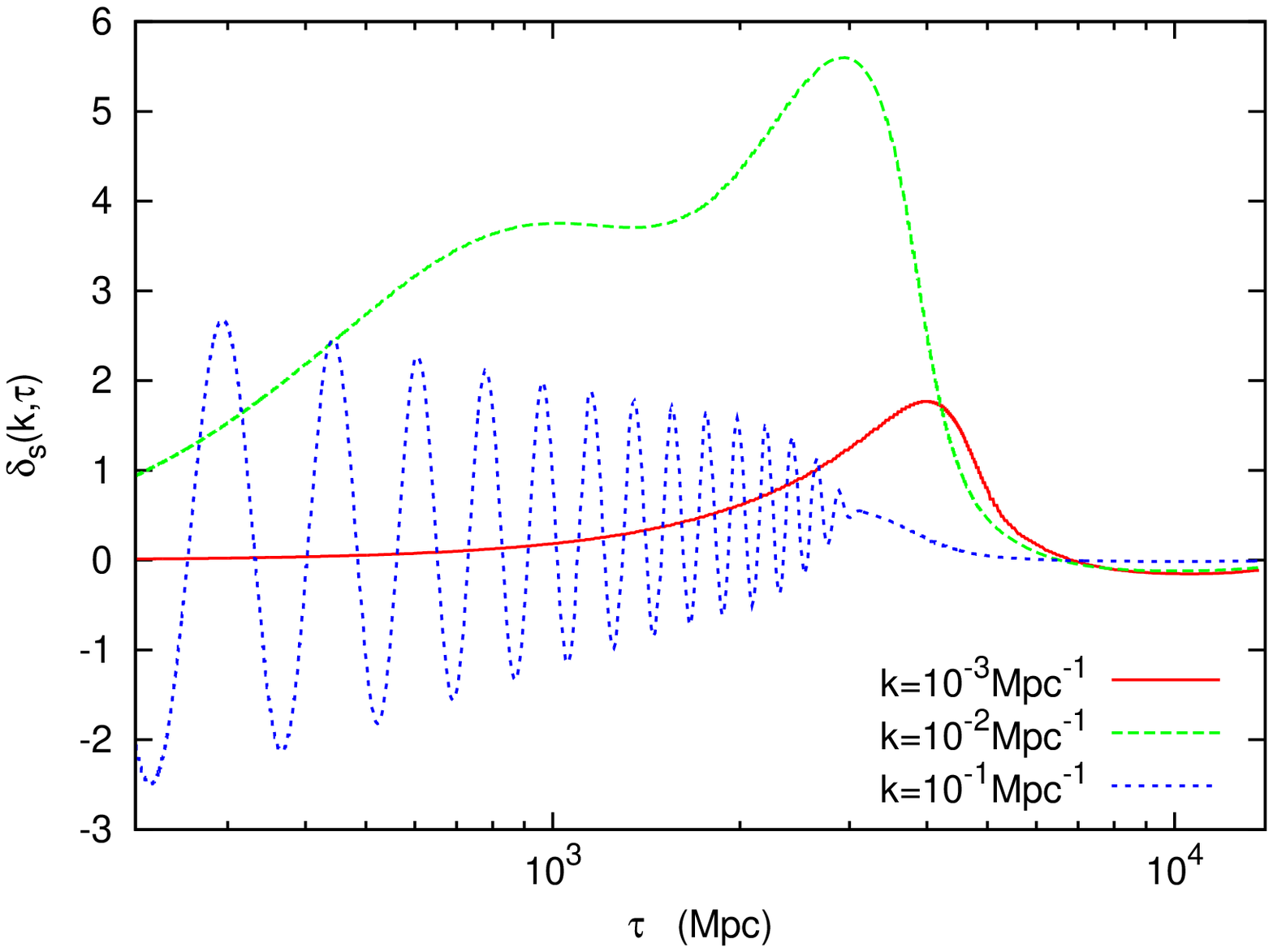}
\end{tabular}
\caption{
Left: Evolution of a few characteristic wavenumbers: that of Hubble crossing,
$k_H=aH$ (bottom black solid line); the Jeans wavenumber (upper blue solid line); and an effective Jeans wavenumber including some effect from viscosity (dashed line; see the text for details). Right: Evolution of the density perturbation $\delta_s$ for three wavenumbers between conformal time $\tau=200$~Mpc and today. These two plots are obtained for the reference model with $(\epsilon, \lambda) = (0.7,0.25)$.}
\label{fig:Perturbation_evolution}
\end{figure}

We first study the evolution of a few characteristic scales related to perturbations in the viscous fluids. For a perfect fluid, the qualitative behavior of perturbations is essentially captured by the Jeans wavenumber:
\begin{equation}
k_J = a \sqrt{\frac{4 \pi G \rho_i}{c_{i}^{2}}} = \sqrt{\frac{3}{2}} \left(\frac{aH}{c_i}\right)~,
\end{equation}
where $c_{i}$ is the sound speed in the fluid. Note that the sound speed is usually defined as
the variation of the pressure with respect to the density for constant entropy. Since a viscous fluid contains entropy perturbations, we cannot readily obtain $c_{s}^{2}$ from the ratio $\delta p_{\rm eff}/\delta \rho_s$.  We should instead bear in mind that the viscosity of the fluid remains negligible in an adiabatic transformation, so we must define the sound speed for the scalar fluid using the equilibrium pressure:
\begin{equation}
c_{s}^{2} = \left. \frac{\partial p_{\rm eff}}{\partial \rho_s} \right|_S
= \frac{\partial p_s}{\partial \rho_s}
= \frac{p_s'(x)}{\rho_s'(x)}~.
\end{equation}
The Jeans wavenumber obtained with $c_{i}=c_{s}$ is shown as a solid blue line in 
figure~\ref{fig:Perturbation_evolution}  (for the same reference model as in the previous section).  For comparison, the dashed line is obtained using
$c_{i}^{2}=P_{\rm eff}(x)/D(x)$, in order to give an indication of when the viscosity becomes important and which scales it can affect.

Let us now summarize the expected behavior of perturbations in the scalar fluid when their wavenumbers
cross the four different regions shown in figure~\ref{fig:Perturbation_evolution} (left).  After crossing the Hubble radius $k=aH$, the modes experience gravitational collapse as long as $k < k_J$.  This stage is very brief: even if the scalar particles become non-relativistic before
the dark-energy-like stage (as it is the case in our reference model), the sound speed decreases very slowly below $1/\sqrt{3}$ and the Jeans wavenumber remains of the same order of magnitude as $aH$. When the modes cross the Jeans length ($k>k_J$), and as long as viscosity is negligible, acoustic waves start to propagate in the fluid with a velocity $c_{s}$. In Fourier space, $\delta_s$ oscillates with a pulsation $k c_{s}$.

At some point, viscosity becomes important, and due to non-adiabatic contributions $|\delta p_{\rm eff}|$ becomes very large with respect to $c_s |\delta \rho_s|$. In our reference model, this happens around $\tau=3000$~Mpc, since this is the time at which the two characteristic scales plotted in figure~\ref{fig:Perturbation_evolution} become different. The very large value of $|\delta p_{\rm eff}/\delta \rho_s|$ implies that the friction term in the equation of propagation of $\delta_s$ drives the perturbation to zero, on a time scale shorter than the period of oscillation. This is the regime in which viscosity erases all perturbations in the fluid.

All these expectations can be checked in figure~\ref{fig:Perturbation_evolution} (right). 
For the reference model, the mode $\delta_s(k=10^{-3}\mathrm{Mpc})$ experiences a very short stage of gravitational amplification and is then driven to zero. The mode $\delta_s(k=10^{-2}\mathrm{Mpc})$ has three visible stages: amplification, oscillation (for just half-a-period) and damping. The mode $\delta_s(k=10^{-1}\mathrm{Mpc})$ is already in the acoustic oscillation phase at the earliest time shown in the figure. Around $\tau=3000$~Mpc, it also experiences a brutal viscosity damping.

The conclusion of this section is that the viscous fluid perturbations never grow significantly during matter domination and are washed out when viscosity becomes important (i.e. prior to dark energy perturbation). From this point of view, our viscous fluid model should be indistinguishable from any other dark energy model with negligible perturbations, like e.g. a scalar field with $c_{s}=1$.

The only time at which the viscous fluid perturbation can play a role is in the radiation domination epoch. During that stage, the fluid dilutes like $\rho_s \propto a^{-4}$ and from the background point of view it can be treated as extra massless neutrino species, as explained in section~\ref{sec:Results_background}. This is not obvious at the perturbation level, since decoupled massless neutrinos are collisionless with anisotropic pressure leading to shear viscosity damping. Instead, our fluid is self-coupled and we assume its anisotropic stress to vanish. Hence, during radiation domination and before the time at which viscosity becomes important, we expect $\delta_s$ to oscillate with a larger amplitude than $\delta_\nu$, even if these two fluid share the same initial conditions and background evolution. This effect could in principle be significant. For instance, it has been proved that the damping of neutrino oscillations caused by shear viscosity is measurable with WMAP data~\cite{DeBernardis:2008ys}. If detectable, this effect would lead to slightly higher CMB peaks in the viscous model than in a wCDM model with the same $N_{\rm eff}$.

\begin{figure}
\centering
\begin{tabular}{c}
\includegraphics[width=0.5\textwidth]{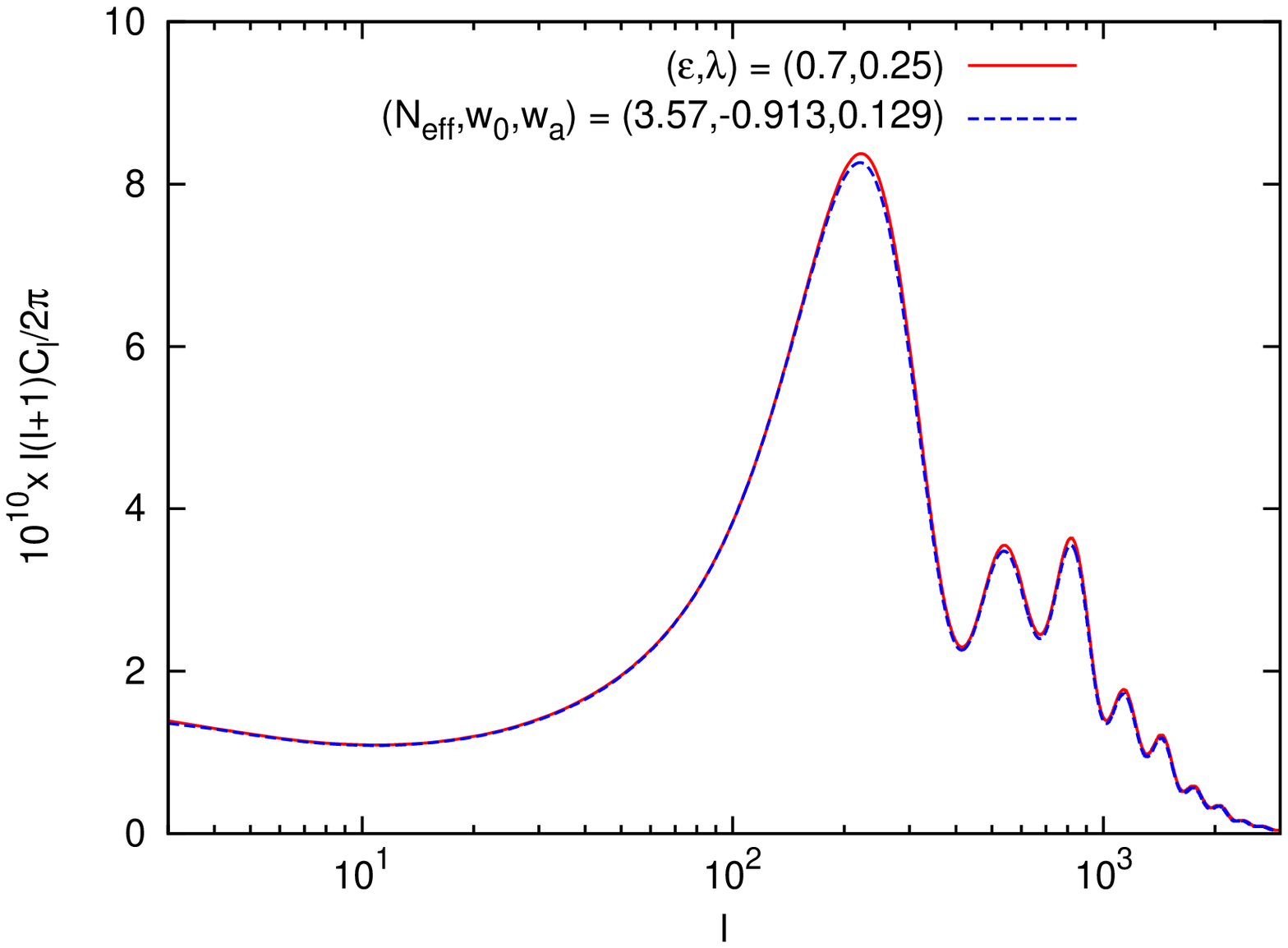}
\includegraphics[width=0.5\textwidth]{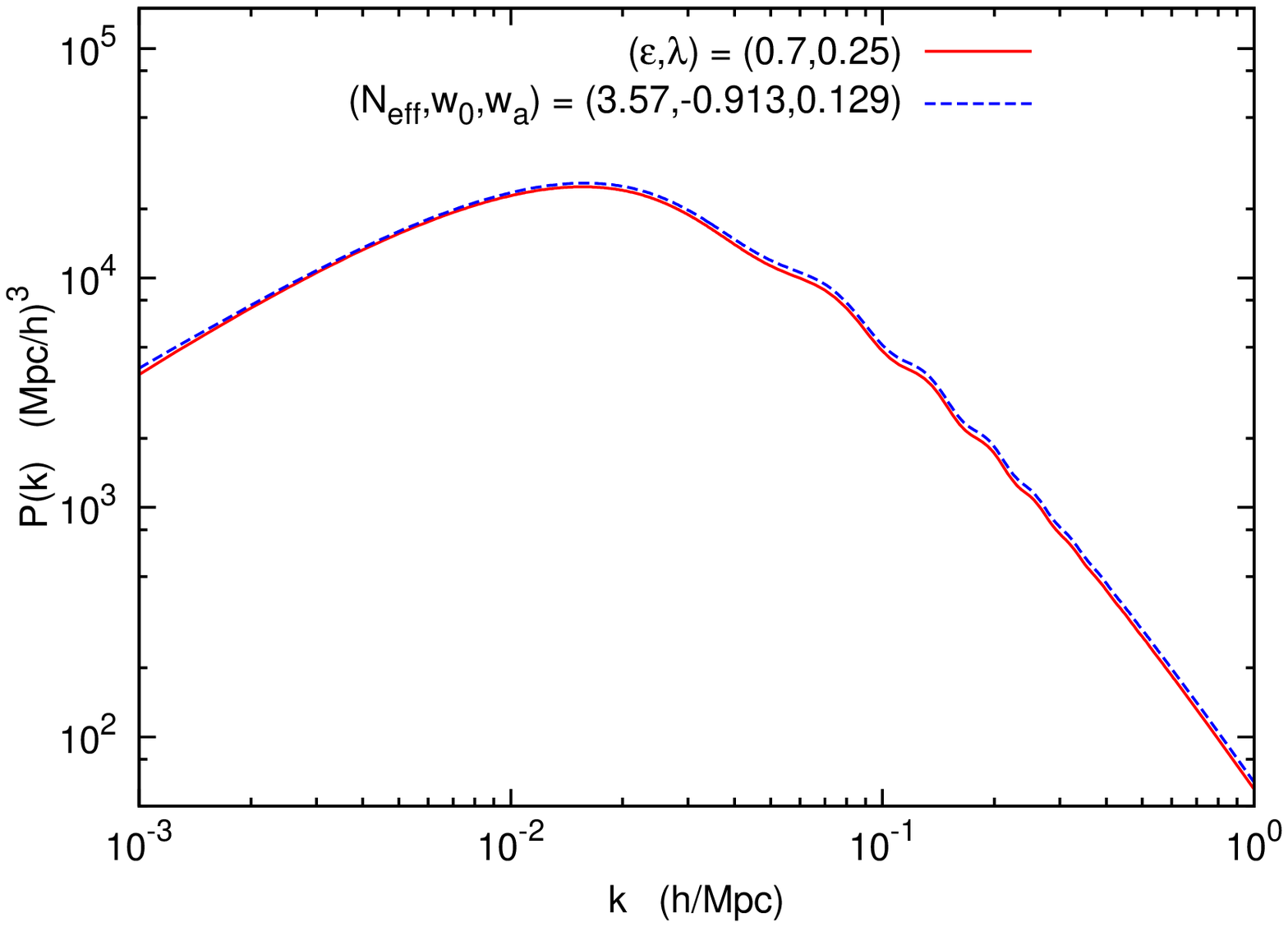}\\
\includegraphics[width=0.5\textwidth]{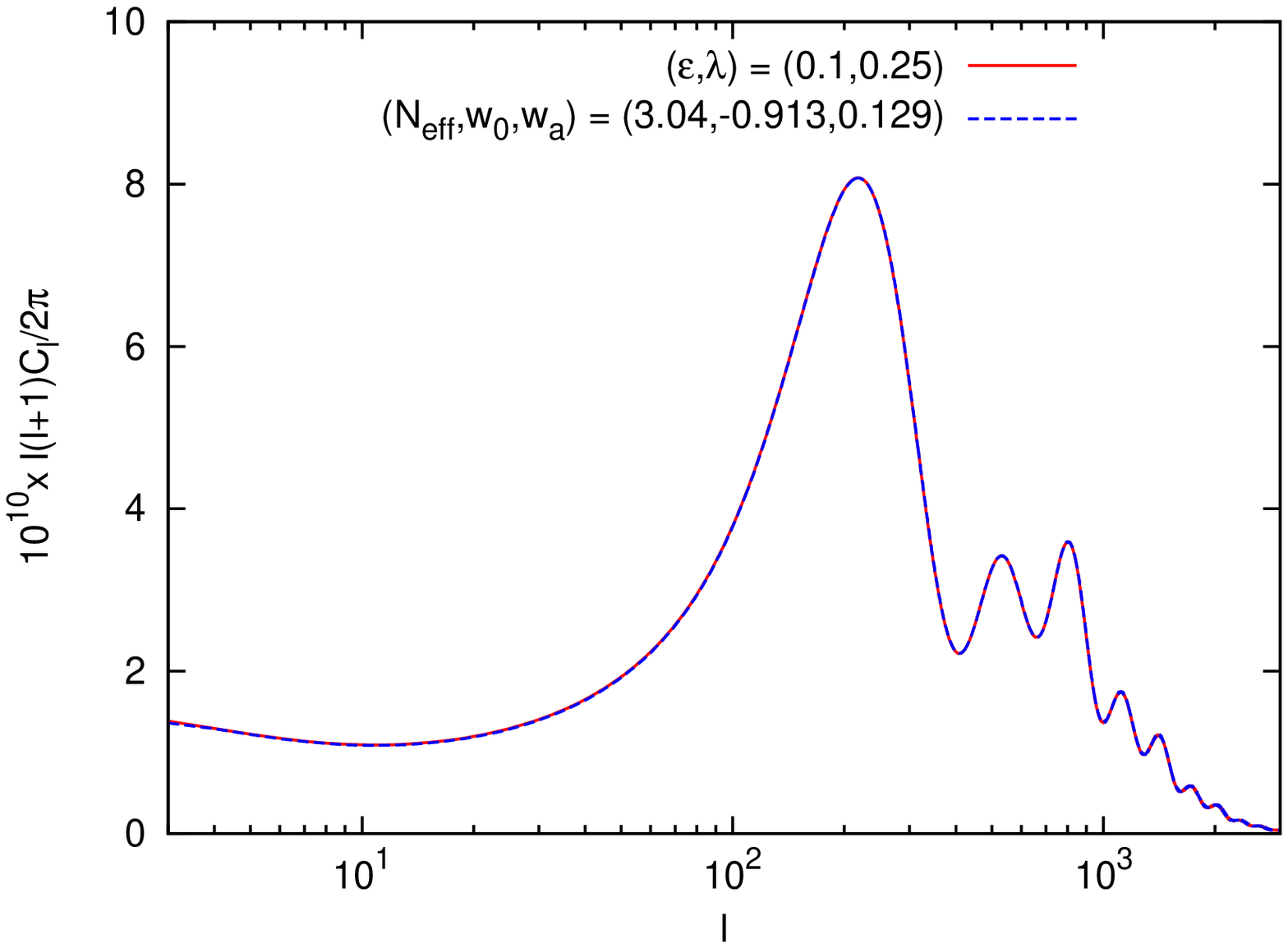}
\includegraphics[width=0.5\textwidth]{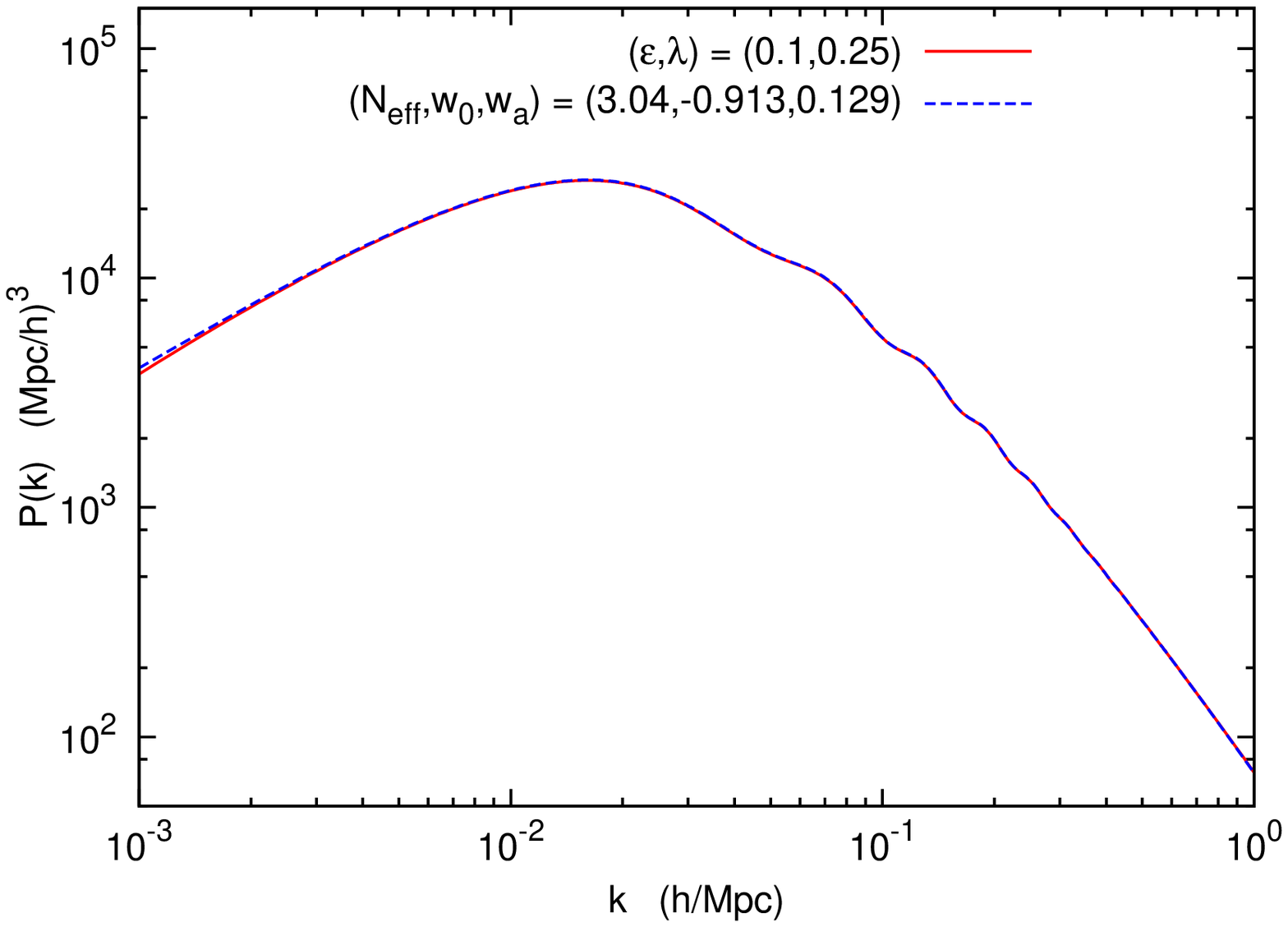}\\
\includegraphics[width=0.5\textwidth]{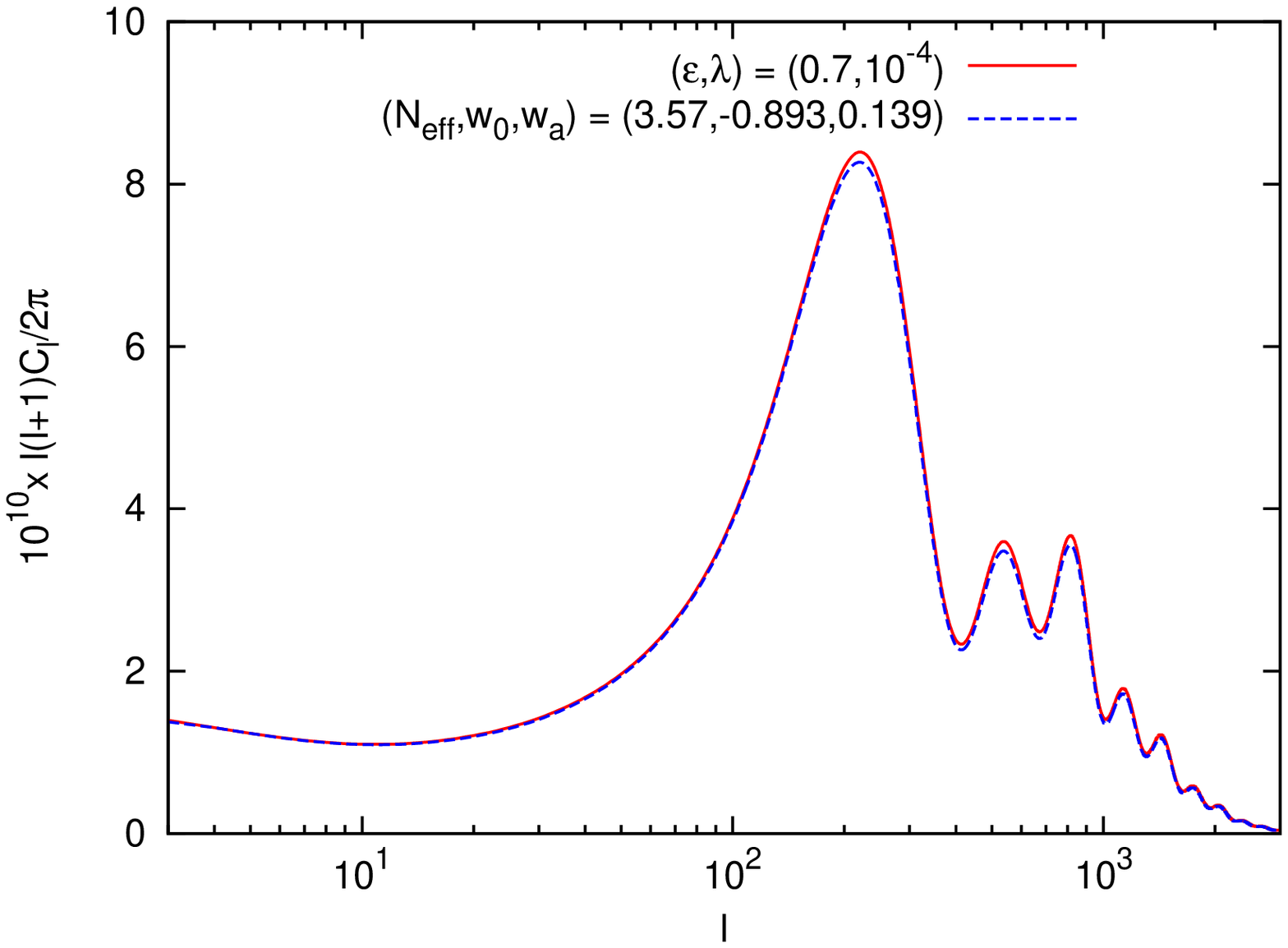}
\includegraphics[width=0.5\textwidth]{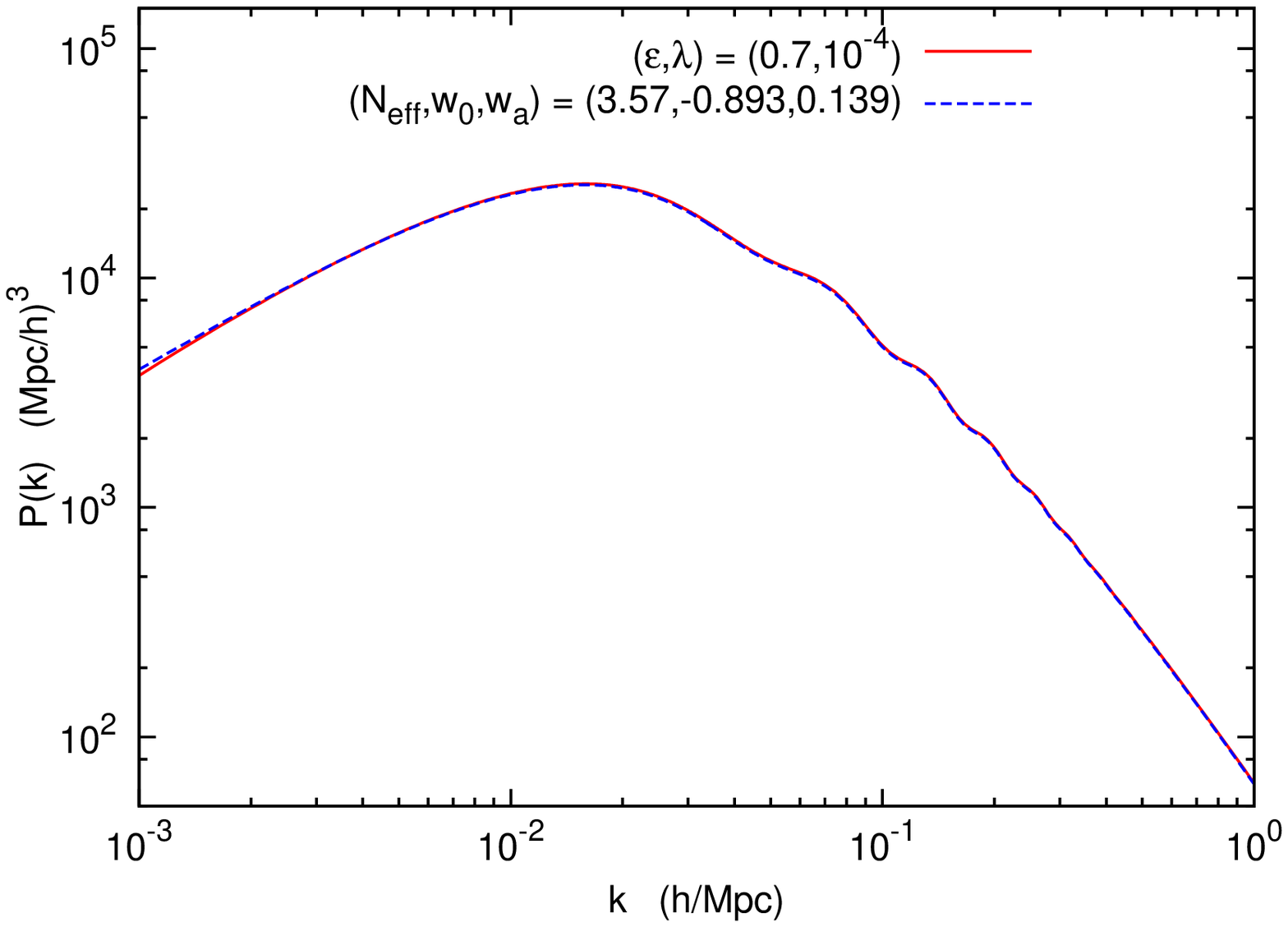}
\end{tabular}
\caption{CMB temperature spectrum (left) and matter power spectrum (right) for pairs of models.  In each plot, we present one viscous fluid model and a nearly equivalent extended wCDM model with the same values of ($N_{\rm eff}$, $w_0$, $w_a$) obtained with the fitting formulas~\ref{eq:fit_neff}--\ref{eq:fit_wa}. From top to bottom, we assume one of the three models already studied in section~\ref{sec:Results_background}.}
\label{fig:spectra}
\end{figure}

These expectations can be checked by running {\tt CLASS} for a pair of models: a viscous fluid model and an extended wCDM model with the same values of ($N_{\rm eff}$,  $w_0$, $w_a$) obtained with the fitting formulas~\ref{eq:fit_neff}--\ref{eq:fit_wa}. For such pairs of models and three choices of parameters (same as in section \ref{sec:Results_background}), we present in figure~\ref{fig:spectra} the CMB temperature and matter power spectra.  For the model with $\epsilon=0.1$ ($N_{\rm eff} \simeq 3.04$), the matching is impressive and confirms our expectations.  For the other two models with 
$\epsilon=0.7$ ($N_{\rm eff} \simeq 3.57$), slightly larger CMB peaks can indeed be observed in the viscous model, for the reason explained above.  This effect is however small, and we recall that $\epsilon=0.7$ is an extreme assumption: it implies that the scalar particles decouple from the thermal bath late enough so that no entropy creation occured afterwards, except at the time of electron-positron annihilation.

\section{Compatibility with observations}
\label{sec:Compatibility_observations}

If the viscous fluid or ``Dark Goo'' model was leading to very specific predictions, we would perform a separate analysis to check its compatibility with observations and would infer confidence limits on each free parameters.  However, we establish in sections~\ref{sec:Parameter_dependence} and \ref{sec:Results_perturbations} that at the level of background and linear perturbation evolution, the minimal Dark Goo model is nearly equivalent to an extended wCDM scenario with the following free parameters beyond those of $\Lambda$CDM:
\begin{itemize}
	\item An effective neutrino number
\begin{equation}
N_{\rm eff}= 3.04 + 2.2 \epsilon^4~.
\end{equation}

	\item A current value of the equation-of-state parameter
\begin{equation}
w_0=-0.9085+0.21(\Omega_m-0.3)+3 \lambda^{-0.4} 10^{-4}~.
\end{equation}

	\item A time variation of the equation-of-state parameter
\begin{equation}
w_a=0.129+0.02(\Omega_m-0.3)+\lambda^{-0.5} 10^{-4}~.
\end{equation}
\end{itemize}
Various analysis have been carried for models with free $N_{\rm eff}$ or $(w_0, w_a)$, but these parameters are usually not explored altogether at the same time. However, in our case the three quantities ($N_{\rm eff}$, $w_0$, $w_a$) are expected to remain close to their
minimal $\Lambda$CDM value (3.04, -1, 0), and by considering bounds obtained separately
on these parameters, we do not expect to make a large error. We also neglect as a first approximation the fact that the viscous model predicts slightly enhanced CMB peaks when $N_{\rm eff}$ is significantly increased (i.e. when $\epsilon>0.5$).

The strongest observational limits on $N_{\rm eff}$ come from Big Bang Nucleosynthesis, $N_{\rm eff} = 3.8 \pm 0.8$ at the 95\% C.L. \cite{Izotov_Thuan_2010}, while a combination of CMB and LSS experiments shows a marginal preference for values $N_{\rm eff}>3$~\cite{WMAP7_2010,Keisler_etal_2011}.  If this preference is confirmed, our model could account for any value of $N_{\rm eff}$ in the range $[3.04, 3.62]$, as long as we assume that the scalar particles decouple before electron-positron annihilation such that $\epsilon \leq T_{\nu}/T_{\gamma}$ (or eventually for $N_{\rm eff}$ in the range $[3.62,5.25]$ if they decouple later such that $\epsilon \leq 1$, but this case would require a specific study of Big Bang Nucleosynthesis and photon recombination, and is not covered by the present paper).  However, current error bars on $N_{\rm eff}$ are large and do not allow to discriminate between models with a late scalar decoupling ($3.04 < N_{\rm eff} <3.62$) and early decoupling ($N_{\rm eff} \simeq 3.04$).

\begin{figure}
\centering
\begin{tabular}{c}
\includegraphics[width=0.5\textwidth]{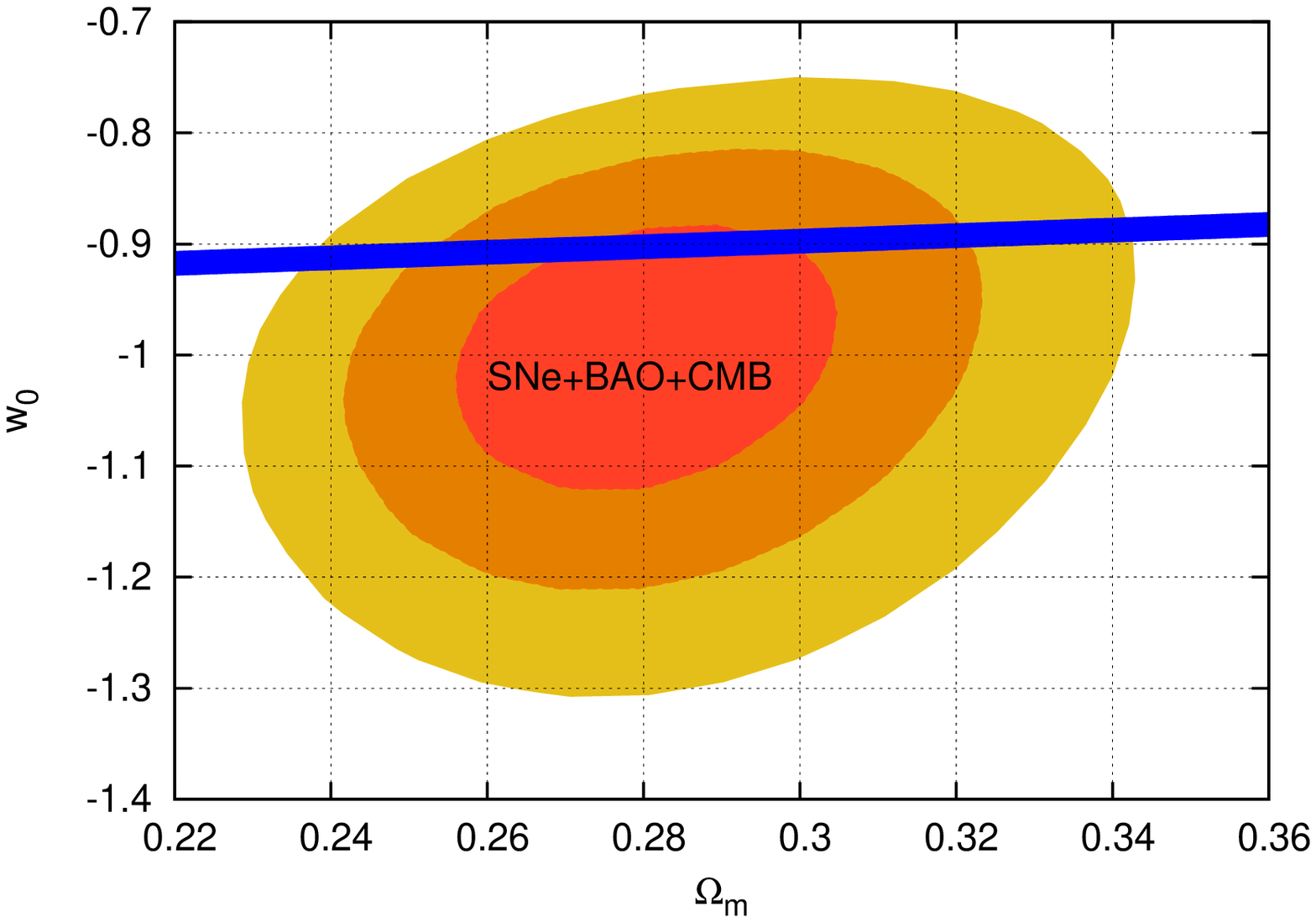}
\includegraphics[width=0.5\textwidth]{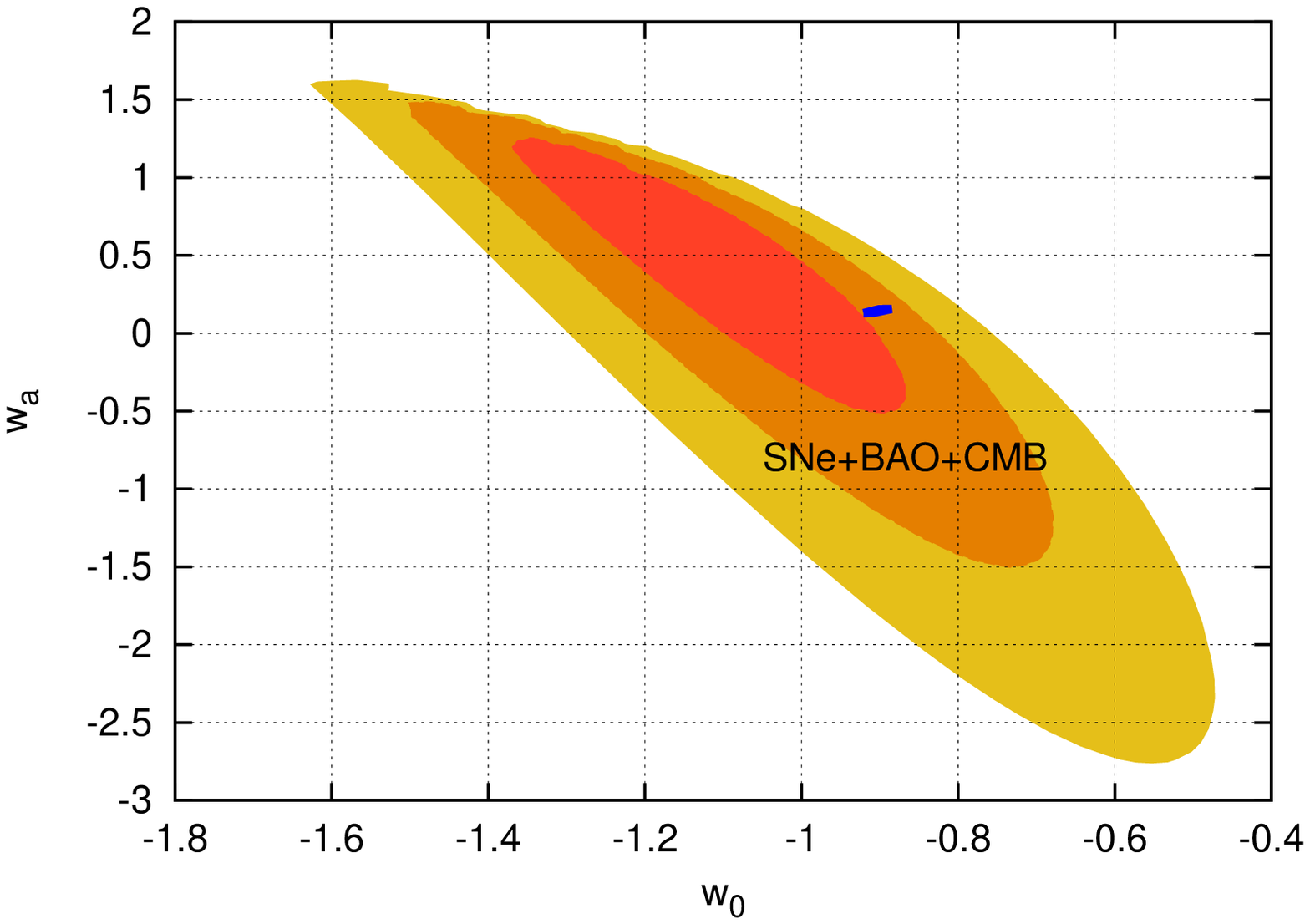}
\end{tabular}
\caption{
Predictions of the Dark Goo model for the parameters $w_0$ and $w_a$, compared to the 68.3\%, 
95.4\% and 99.7\% confidence regions obtained from a joint fit of Supernovae, BAO and CMB data
 \cite{Amanullah:2010vv}.  Left: Prediction for $w_0$ for each given $\Omega_m$, when the other 
parameters of the model are varied in the following ranges: $h \in [0.714,0.724]$, and $\lambda \in [2\times10^{-5},1]$ (corresponding to $m_0 \in [10^{-2}, 2]$~eV).  Right: Predictions for $w_0$ and $w_a$ when other parameters are varied in the ranges described above and $\Omega_m \in [0.245, 0.306]$.}
\label{fig:Predictions_for_w}
\end{figure}

Neglecting the effect of a possibly enhanced value of $N_{\rm eff}$, we can readily use the bounds derived in the literature on the parameters ($\Omega_m$, $w_0$, $w_a$) in order to test the compatibility of our model with current data. We first consider the region in ($\Omega_m$, $w_0$) space preferred by a combination of Supernovae, BAO and CMB data presented in ref.~\cite{Amanullah:2010vv}.
If we assume $h \in [0.714,0.724]$ (as indicated by WMAP+BAO) and $\lambda \in [2\times10^{-5},1]$
(the range of validity of our model), we find that  ($\Omega_m$, $w_0$) must lie inside the thin blue band shown in the left panel of figure~\ref{fig:Predictions_for_w}. Note that this band is obtained numerically with {\tt CLASS}, but using our analytical fitting formula \ref{eq:fit_w0} would be equivalent.
This band has a large overlap with the 68\% preferred region of ref.~\cite{Amanullah:2010vv}, thus the usual bounds on $\Omega_m$ also apply in our case.  This comparison does not allow to discriminate between different values of $\lambda$, since the impact of varying $\lambda$ on $w_0$ is very small compared to the sensitivity of the data to $w_0$. 

Finally, we vary ($h$, $\lambda$) in the same ranges with $\Omega_m$ in his preferred range $[0.245, 0.306]$ and obtain an allowed region in ($w_0$, $w_a$) space.  This region is very small and reduces almost to a point $(-0.9, 0.13)$ with respect to the region preferred by the data, as can be checked on the right panel of figure~\ref{fig:Predictions_for_w}.  This point lies in the 95\% preferred region, very close to the 68\% region, showing that our model is still compatible with current data.

In the future, with better bounds on $N_{\rm eff}$, $w_0$ and $w_a$, it will be possible to better test and maybe to exclude our minimal Dark Goo model, especially if the values $(w_0, w_a)\sim (-0.9,0.13)$ can be ruled out at some point.

\section{Conclusion}
\label{sec:Conclusion}

In summary we present a model of cosmological evolution with an additional scalar field.  We show that in a certain range of values for the mass and self-coupling of the scalar field, bulk viscosity plays an important role at late times and mimics a dark energy behavior.  At the background level, the model is compatible with current data on the equation-of-state parameter and predicts a small time variation for this parameter.  At the perturbation level, the bulk viscous model produces the same temperature anisotropies and matter power spectrum as an extended wCDM model, except in the extreme case where the scalar field decouples from the photons at a very late stage and small deviations in the CMB peak heigths are observed.  The model can also account for the extra relativistic degrees of freedom that are marginally preferred by the data.

The Dark Goo model has several features worth mentioning.  For instance, all parameter ranges are bounded, making the model easily falsifiable by future dark energy experiments (note that the upper bound comes from the use of perturbation theory and is not a strict bound in that sense).  The possible values for the model parameters are ``reasonable'' and there is no apparent fine tuning.  On a more philosophical level, it is also reassuring to know that the model has a built-in mechanism (i.e. breakdown of hydrodynamics) that may prevent the universe from accelerating forever (no ``Big Rip'').  The model is also microscopic and the functional form for the bulk viscosity is obtained from first principles.  This allows for a discussion of the validity of the model in terms of microscopic quantities (coupling, mean free path, etc).  This is to be contrasted with other bulk viscous or Chaplygin gas models, where an exotic equation of state with little or no physical justification is postulated.  


We stress that the whole model hinges on the validity of hydrodynamics.  The assumptions of hydrodynamics have to be satisfied in order for the concept of bulk viscosity to make sense.  We discuss these issues and provide plausibility arguments for each assumptions, but these arguments can certainly be improved.  In particular, a more fundamental understanding of cavitation would be helpful.




\begin{acknowledgments}
The authors would like to thank S. Jeon, P. Huovinen, M. Shaposhnikov, J. Berges, J. Cline and R. 
Brandenberger for useful comments and discussions.  J.-S. G. would like to thank the Laboratoire de Physique des Particules et Cosmologie at EPFL for its kind hospitality and for providing the facilities necessary for the completion of this work.
\end{acknowledgments}

\appendix

\section{Modifications to perturbation equations in \texorpdfstring{\tt CLASS}{CLASS}} 
\label{sec:appendix}

In section~\ref{sec:Linear_perturbations}, we present the perturbation equations for the viscous fluid using the conventions of ref.~\cite{Weinberg_2008}.  In order to make our results easily reproducible, we would like to recapitulate in this appendix the list of changes implemented in the {\tt CLASS} code in order to account for these perturbations. {\tt CLASS} uses the notations of ref.~\cite{Ma:1995ey}.
Density perturbations are represented by $\delta_i$'s and velocity divergences by $\theta_i$'s.
We implement the new equations in the usual synchronous gauge comoving with cold dark matter 
($\theta_{cdm}=0$), with metric perturbations $\eta$ and $h$. The code uses conformal time $\tau$, and throughout this appendix the dots denote derivatives with respect to $\tau$, unlike in section \ref{sec:Linear_perturbations}. We define $\mathcal{H}=\dot{a}/a=aH$.

When the system of perturbed equations is integrated over time for each wavenumber,
at each new step we interpolate in the table of background quantities in order to obtain $\rho_s$, $p_s$, $\zeta$, $p_{\rm eff}=p_s-3H\zeta$, $D\equiv\rho_s'$, and finally
\begin{equation}
P_{\rm eff}   
\equiv
\frac{\partial p_{\rm eff}}{\partial x} 
=
\frac{m_0^4}{2 \pi^2}
\left( 
\frac{K_2'(x)}{x^2}
-2 \frac{K_2(x)}{x^3} \right)
- 3 H \left( \kappa_2 \frac{m_{\rm th}}{m_0}+\frac{\kappa_3}{x}\right) \zeta~.
\end{equation}
We then infer the quantities $w_s=p_{\rm eff}/\rho_s$ and
\begin{equation}
 \dot{w}_s =  3 \mathcal{H} (1+w_s) (-P_{\rm eff}   /D+w_s)
 - 3 \dot{H} \zeta / \rho_s,
\end{equation}
(note that there is no typo here: this formula mixes $\mathcal{H}$ and $H$). 
The effective pressure perturbation $\delta p_{\rm eff}$ is derived from
\begin{equation}
 \delta p_{\rm eff} = \frac{P_{\rm eff}}{D} \rho_s \delta_s
 -a^{-1} \left( \theta_s+\frac{\dot{h}}{2} \right) \zeta~,
\end{equation}
which is equivalent to eq.~\ref{eq:Perturbation_effective_pressure}, replacing $\delta x$ by $\rho_s \delta_s /D$, and with the identification $\psi \equiv (dh/dt)/2 = a^{-1} \dot{h}/2$ and $a^{-1} \nabla^2 \delta U \equiv \theta_s$.  The continuity and Euler equations for the viscous scalar fluid read:
\begin{eqnarray}
\dot{\delta}_s &=& -(1+w_s)\left(\theta_s + \frac{\dot{h}}{2}\right)-3 \mathcal{H}
\left(\frac{\delta p_{\rm eff}}{\rho_s}-w_s\delta_s\right), \\
\dot{\theta}_s &=& - \mathcal{H} (1-3w_s) \theta_s - \frac{\dot{w}_s}{1+w_s}\theta_s
+\frac{k^2 \, \delta p_{\rm eff}}{(1+w_s)\rho_s}~.
\end{eqnarray}
In the Einstein equations, we include $\rho_s \delta_s$ and $(\rho_s+p_{\rm eff}) \theta_s$
in the calculation of the total $\delta \rho$ and $(\rho+p) \theta$ source terms.

Finally, as far as initial conditions are concerned, we treat the viscous scalar fluid at very early times as any other extra relativistic degrees of freedom. There is actually a difference between massless neutrinos and our fluid at early times:
the latter are collisionless, while the former are self-coupled. This implies that neutrinos can have a significant anisotropic stress, while we assume that the scalar fluid can maintain an isotropic pressure. However, on super-Hubble scales on which initial conditions are implemented, this difference should have a negligible impact.

Usually, the expression of initial conditions for all species involves
a parameter $R_{\nu}$ defined as the ratio of the massless neutrino density over the total ultra-relativistic matter densities $R_{\nu}=\rho_\nu/\rho_r$.
We include $\rho_s$ in both the numerator and denominator. We then derive initial conditions for all species in the usual way, including the density and velocity perturbations of ultra-relativistic species
($\delta_{\tt ur}$, $\theta_{\tt ur}$). We initialize ($\delta_s$, $\theta_s$) to precisely the same values
as ($\delta_{\tt ur}$, $\theta_{\tt ur}$).

\bibliographystyle{unsrtnat}
\bibliography{paper_bulk_viscosity_bibliography}

\end{document}